\documentclass[
 superscriptaddress,
 reprint, 
 amsmath,amssymb,
 aps,nofootinbib
]{revtex4-2}
\usepackage[utf8]{inputenc}
\usepackage{graphics}
\usepackage{graphicx}
\usepackage{dcolumn}
\usepackage{bm}
\usepackage{braket}
\usepackage{textcomp}
\usepackage{natbib}
\usepackage[dvipsnames]{xcolor}
\usepackage[colorlinks]{hyperref} 
\hypersetup{
      colorlinks=true,
      linkcolor=blue,
      filecolor=blue,
      citecolor = teal,      
      urlcolor=blue,
      }
\begin{document}
\title{Quasiprojectile breakup and isospin equilibration at Fermi energies:\\ an indication of longer projectile-target contact times?}

\author{C.~Ciampi}
 \email{caterina.ciampi@ganil.fr}
 \affiliation{Grand Accélérateur National d'Ions Lourds (GANIL), CEA/DRF–CNRS/IN2P3, Boulevard Henri Becquerel, F-14076 Caen, France}
\author{S.~Piantelli}
 \affiliation{INFN - Sezione di Firenze, 50019 Sesto Fiorentino, Italy}
\author{G.~Casini}
 \affiliation{INFN - Sezione di Firenze, 50019 Sesto Fiorentino, Italy}
\author{A.~Ono}
 \affiliation{Department of Physics, Tohoku University, Sendai 980-8578, Japan}
\author{J.~D.~Frankland}
 \affiliation{Grand Accélérateur National d'Ions Lourds (GANIL), CEA/DRF–CNRS/IN2P3, Boulevard Henri Becquerel, F-14076 Caen, France}
\author{L.~Baldesi}
\affiliation{INFN - Sezione di Firenze, 50019 Sesto Fiorentino, Italy}
\affiliation{Dipartimento di Fisica e Astronomia, Universit\`{a} di Firenze, 50019 Sesto Fiorentino, Italy}
\author{S.~Barlini}
\affiliation{INFN - Sezione di Firenze, 50019 Sesto Fiorentino, Italy}
\affiliation{Dipartimento di Fisica e Astronomia, Universit\`{a} di Firenze, 50019 Sesto Fiorentino, Italy}
\author{B.~Borderie}
\affiliation{Université Paris-Saclay, CNRS/IN2P3, IJCLab, 91405 Orsay, France}
\author{R.~Bougault}
 \affiliation{Université de Caen Normandie, ENSICAEN, CNRS/IN2P3, LPC Caen UMR6534, F-14000 Caen, France}
\author{A.~Camaiani}
 \affiliation{INFN - Sezione di Firenze, 50019 Sesto Fiorentino, Italy}
 \affiliation{Dipartimento di Fisica e Astronomia, Universit\`{a} di Firenze, 50019 Sesto Fiorentino, Italy}
\author{A.~Chbihi}
\affiliation{Grand Accélérateur National d'Ions Lourds (GANIL), CEA/DRF–CNRS/IN2P3, Boulevard Henri Becquerel, F-14076 Caen, France}
\author{J.~A.~Dueñas}
 \affiliation{Departamento de Ingeniería Eléctrica y Centro de Estudios Avanzados en Física, Matemáticas y Computación, Universidad de Huelva, 21007 Huelva, Spain}
\author{Q.~Fable}
\affiliation{Laboratoire des 2 Infinis - Toulouse (L2IT-IN2P3), Universit\'e de Toulouse, CNRS, UPS, F-31062 Toulouse Cedex 9 (France)}
\author{D.~Fabris}
\affiliation{INFN - Sezione di Padova, 35131 Padova, Italy}
\author{C.~Frosin}
\affiliation{INFN - Sezione di Firenze, 50019 Sesto Fiorentino, Italy}
\affiliation{Dipartimento di Fisica e Astronomia, Universit\`{a} di Firenze, 50019 Sesto Fiorentino, Italy}
\author{T.~Génard}
\affiliation{Grand Accélérateur National d'Ions Lourds (GANIL), CEA/DRF–CNRS/IN2P3, Boulevard Henri Becquerel, F-14076 Caen, France}
\author{F.~Gramegna}
\affiliation{INFN - Laboratori Nazionali di Legnaro, 35020 Legnaro, Italy}
\author{D.~Gruyer}
\affiliation{Université de Caen Normandie, ENSICAEN, CNRS/IN2P3, LPC Caen UMR6534, F-14000 Caen, France}
\author{M.~Henri}
\affiliation{Grand Accélérateur National d'Ions Lourds (GANIL), CEA/DRF–CNRS/IN2P3, Boulevard Henri Becquerel, F-14076 Caen, France}
\author{B.~Hong}
\affiliation{Center for Extreme Nuclear Matters (CENuM), Korea University, Seoul 02841, Republic of Korea}
\affiliation{Department of Physics, Korea University, Seoul 02841, Republic of Korea}
\author{S.~Kim}
 \affiliation{Center for Exotic Nuclear Studies, Institute for Basic Science, Daejeon 34126, Republic of Korea}
\author{A.~Kordyasz}
\affiliation{Heavy Ion Laboratory, University of Warsaw, 02-093 Warszawa, Poland}
\author{T.~Kozik}
\affiliation{Marian Smoluchowski Institute of Physics Jagiellonian University, 30-348 Krakow, Poland}
\author{M.~J.~Kweon}
\affiliation{Department of Physics, Inha University, Incheon 22212, Republic of Korea}
\author{N.~Le~Neindre}
 \affiliation{Université de Caen Normandie, ENSICAEN, CNRS/IN2P3, LPC Caen UMR6534, F-14000 Caen, France}
\author{I.~Lombardo}
\affiliation{INFN - Sezione di Catania, 95123 Catania, Italy}
\affiliation{Dipartimento di Fisica e Astronomia, Universit\`a di Catania, via S. Sofia 64, 95123 Catania, Italy}
\author{O.~Lopez}
\affiliation{Université de Caen Normandie, ENSICAEN, CNRS/IN2P3, LPC Caen UMR6534, F-14000 Caen, France}
\author{T.~Marchi}
 \affiliation{INFN - Laboratori Nazionali di Legnaro, 35020 Legnaro, Italy}
\author{K.~Mazurek}
 \affiliation{Institute of Nuclear Physics Polish Academy of Sciences, PL-31342 Krakow, Poland}
\author{S.~H.~Nam}
\affiliation{Center for Extreme Nuclear Matters (CENuM), Korea University, Seoul 02841, Republic of Korea}
\affiliation{Department of Physics, Korea University, Seoul 02841, Republic of Korea}
\author{J.~Park}
\affiliation{Center for Extreme Nuclear Matters (CENuM), Korea University, Seoul 02841, Republic of Korea}
\affiliation{Department of Physics, Korea University, Seoul 02841, Republic of Korea}
\author{M.~P\^{a}rlog}
\affiliation{Université de Caen Normandie, ENSICAEN, CNRS/IN2P3, LPC Caen UMR6534, F-14000 Caen, France}
\affiliation{``Horia Hulubei'' National Institute for R\&D in Physics and Nuclear Engineering (IFIN-HH), P.~O.~Box MG-6, Bucharest Magurele, Romania}
\author{G.~Pasquali}
 \affiliation{INFN - Sezione di Firenze, 50019 Sesto Fiorentino, Italy}
 \affiliation{Dipartimento di Fisica e Astronomia, Universit\`{a} di Firenze, 50019 Sesto Fiorentino, Italy}
\author{G.~Poggi}
 \affiliation{INFN - Sezione di Firenze, 50019 Sesto Fiorentino, Italy}
 \affiliation{Dipartimento di Fisica e Astronomia, Universit\`{a} di Firenze, 50019 Sesto Fiorentino, Italy}
\author{A.~Rebillard-Soulié}
\affiliation{Université de Caen Normandie, ENSICAEN, CNRS/IN2P3, LPC Caen UMR6534, F-14000 Caen, France}
\author{A.~A.~Stefanini}
\affiliation{INFN - Sezione di Firenze, 50019 Sesto Fiorentino, Italy}
\affiliation{Dipartimento di Fisica e Astronomia, Universit\`{a} di Firenze, 50019 Sesto Fiorentino, Italy}
\author{S.~Upadhyaya}
 \affiliation{Marian Smoluchowski Institute of Physics Jagiellonian University, 30-348 Krakow, Poland}
\author{S.~Valdré}
\affiliation{INFN - Sezione di Firenze, 50019 Sesto Fiorentino, Italy}
\author{G.~Verde}
\affiliation{INFN - Sezione di Catania, 95123 Catania, Italy}
\affiliation{Laboratoire des 2 Infinis - Toulouse (L2IT-IN2P3), Universit\'e de Toulouse, CNRS, UPS, F-31062 Toulouse Cedex 9 (France)}
\author{E.~Vient}
\affiliation{Université de Caen Normandie, ENSICAEN, CNRS/IN2P3, LPC Caen UMR6534, F-14000 Caen, France}
\author{M.~Vigilante}
\affiliation{INFN - Sezione di Napoli, 80126 Napoli, Italy}
\affiliation{Dipartimento di Fisica, Università di Napoli, 80126 Napoli, Italy}

\collaboration{INDRA-FAZIA collaboration}
\begin{abstract}
 An investigation of the quasiprojectile breakup channel in semiperipheral and peripheral collisions of $^{58,64}$Ni+$^{58,64}$Ni at 32 and 52~MeV/nucleon is presented. Data have been acquired in the first experimental campaign of the INDRA-FAZIA apparatus in GANIL. The effect of isospin diffusion between projectile and target in the two asymmetric reactions has been highlighted by means of the isospin transport ratio technique, exploiting the neutron-to-proton ratio of the quasiprojectile reconstructed from the two breakup fragments.
 We found evidence that, for the same reaction centrality, a higher degree of relaxation of the initial isospin imbalance is achieved in the breakup channel with respect to the more populated binary output, possibly indicating the indirect selection of specific dynamical features.
 We have proposed an interpretation based on different average projectile-target contact times related to the two exit channels under investigation, with a longer interaction for the breakup channel. The time information has been extracted from AMD simulations of the studied systems coupled to \textsc{Gemini++}: the model calculations support the hypothesis hereby presented.
\end{abstract}

\maketitle

\section{Introduction}
The effort to investigate the properties of nuclear matter far from equilibrium conditions has been going on for a few decades \cite{Li2013, Huth2022}.
More specifically, there is a great interest in constraining the values of the parameters defining the density dependence of the symmetry energy $E_{sym}(\rho)$ of the nuclear equation of state (NEoS), which has implications in both nuclear physics and astrophysics and nowadays still presents some issues \cite{Reed2021}. The rich phenomenology associated to heavy ion collisions provides multiple tools to gather information on this topic, e.g., the study of isospin transport phenomena \cite{Baran2005}. Two main contributions to such nucleon exchange processes are generally distinguished: the \textit{isospin diffusion}, taking place whenever an isospin gradient is present and inducing the isospin equilibration, and the \textit{isospin drift}, which is driven by a density gradient, leading to the neutron enrichment of lower-density regions.

In a previous paper \cite{Ciampi2022} the topic of isospin equilibration in nuclear reactions at Fermi energies (20-100~MeV/nucleon) has been addressed by considering the most abundant reaction channel for semiperipheral and peripheral Ni-Ni reactions. Indeed the outcome of heavy ion reactions in the Fermi energy regime can vary substantially depending on the centrality of the collision.
For semiperipheral and peripheral reactions, the binary exit channel is the dominant one, resulting in the production of two main fragments, the quasiprojectile (QP) and the quasitarget (QT). Together with the QP and QT, lighter ejectiles, such as neutrons, light charged particles (LCPs, $Z=1,2$) or intermediate mass fragments (IMFs, in this work $Z=3,4$), are also produced in different phases of the process, with a dynamical or statistical origin. The IMFs, being less likely to be produced in the deexcitation of the QP/QT, are mostly emitted in the midvelocity region: this is generally interpreted as the result of the rupture of the \textit{neck}, i.e., the elongated area connecting the QP and the QT in a late stage of the contact phase \cite{DiToro2006}. The experimentally observed neutron enrichment of midvelocity products is in fact commonly interpreted as an evidence of the action of isospin drift towards such low density region.

Along with the binary exit channel, a ternary (or quaternary) outcome is also possible: in this case, one (or both) of the heavy products breaks up into two smaller fragments at some point of the reaction. Depending on the timescale in which this fission takes place, its physical origin may be different. A distinction is generally made between \textit{statistical fission} and \textit{dynamical fission} (or \textit{breakup}) of the QP/QT. The former process represents one of the possible statistical deexcitation mechanisms for a thermally equilibrated excited fragment produced in the dynamical phase of the reaction, in competition with the evaporation of nucleons and clusters. On the other hand, the fast breakup process \cite{Glassel1983,Casini1993, Stefanini1995} has been linked to a dynamical origin, with shorter characteristic timescales (around or below 1~zs). This fast breakup can be seen as a smooth evolution of the IMF production from midvelocity towards heavier nuclear species produced in such fission process. It also features a strongly anisotropic emission pattern in the reaction plane \cite{Casini1993,Stefanini1995,Piantelli2020,Bocage2000,DeFilippo2005,Jedele2017}: in fact, a larger mass asymmetry between the two breakup products is associated with a configuration more aligned with the QP-QT axis, with the lighter fragment (LF) mostly emitted towards the reaction center-of-mass (c.m.), with respect to the heavy fragment (HF).
Due to kinematics and experimental limitations, the LF and the HF originating from a QP breakup are usually the only ones that can be detected and studied, and therefore also in this work the QT breakup will not be considered and we will refer only to the QP breakup.
According to a possible interpretation of the phenomenon, based on a semiclassical picture and applicable to semiperipheral reactions, the QP and the QT emerging from the collision can be strongly deformed along the separation axis. Their deformation and angular momentum can lead to a prompt breakup of, e.g., the QP in the above mentioned two fragments, with a generally asymmetric split. 
In fact, due to angular momentum conservation, the rotation of the interacting initial di-nuclear system is inherited by the emerging deformed QP/QT fragments; therefore, an $\alpha$ angle between the QP-QT separation axis and the QP fission axis (also referred to as proximity angle $\theta_\text{PROX}$ in some works \cite{Bocage2000,DeFilippo2005}) develops and increases with the time elapsing between the two steps.
The short expected time interval between the QP-QT separation and the QP breakup could explain the preferential backward emission of its LF, that originates from a region closer to the neck. In this respect, as said, the observation of an IMF in the midvelocity region can be interpreted as the most asymmetric case of QP/QT breakup.
The study of the breakup channel can be of interest also in the framework of isospin transport. In fact, due to the short characteristic timescales that might not allow for a complete isospin equilibration between the two breakup fragments, the LF can be expected to keep memory of the previous neutron enrichment at midvelocity.
In this scenario, $\alpha$ has been proposed as a ``clock'' for an estimation of the time scale of the breakup, provided that such time is short enough compared to the period of the QP rotation. At the same time also the degree of isospin equilibration inside the original deformed QP increases, and this should be reflected in the neutron content of the resulting HF and LF. Following this assumption, in Ref.~\cite{Jedele2017} a time scale of the isospin equilibration process has been obtained ($\sim10^{-1}\,$zs); the possible sensitivity of such equilibration time scale to the asy-stiffness of the NEoS is investigated in Ref.~\cite{Jedele2023}. 
However in Ref.~\cite{Piantelli2020} the scenario of a tight correlation between the $\alpha$ angle and the time elapsed from the QP-QT separation to the QP breakup does not seem to be fully supported in the framework of the antisymmetrized molecular dynamics (AMD) model \cite{Ono1992}.
The topics related to the QP breakup are clearly still an open field, and they deserve a deeper investigation.

In this paper we focus on the analysis of the isospin diffusion mechanism between asymmetric projectile and target in events featuring a breakup of the QP. After an overview of the experimental apparatus (Sec.~\ref{sec:experiment}), the selection criterion used for this exclusive analysis is presented in Sec.~\ref{ssec:breakup_selection}. In Sec.~\ref{ssec:isospin_analysis} the effect of relaxation of the initial projectile-target isospin imbalance is highlighted exploiting the isospin transport ratio method \cite{Rami2000} applied to the neutron-to-proton ratio of the QP reconstructed from the two fission fragments. In this respect, in Sec.~\ref{ssec:QPr-QPb_comparison} we evidence some interesting differences between the QP breakup channel and the more populated binary output of the reaction, that we previously studied in Ref.~\cite{Ciampi2022} for the same dataset. Lastly, we propose an interpretation of such observation based on the contact time between projectile and target, which seems to be supported by the AMD calculations (Sec.~\ref{ssec:contact_time}).

\section{The experiment}
\label{sec:experiment}
The INDRA-FAZIA setup, operating in GANIL (Caen, France), has been already described in Refs.~\cite{Lopez2018,Ciampi2022,Casini2022}; here we briefly recall the main features. INDRA \cite{Pouthas1995,Pouthas1996} and FAZIA \cite{Bougault2014,Valdre2019} are both multi-detector apparatuses specifically designed and optimized for the detection and identification of nuclear fragments produced in heavy ion collisions at Fermi energies, though with characteristics that are somewhat complementary. 
Twelve FAZIA blocks (corresponding to 192 detection units), mounted in a wall configuration \cite{Ciampi2022}, cover the most forward polar angles (1.4\textdegree$<\theta<$12.6\textdegree), in order to collect and identify the fragments belonging to the QP phase space, including the heavy QP remnant itself, by exploiting the optimal isotopic identification performance that the FAZIA array is able to provide through different techniques. The high granularity of FAZIA makes it suitable to study the isospin content of both QP breakup fragments, which in most cases can be simultaneously detected and mass identified \cite{Piantelli2020,Camaiani2021}.
On the other hand, twelve INDRA rings (for a total of 240 detection units) cover the polar angles 14\textdegree$<\theta<$176\textdegree; the large angular coverage is useful for a good reconstruction of the global event.
INDRA and FAZIA acquisitions are coupled by means of an event timestamp distribution system based on the GANIL VXI CENTRUM module \cite{CENTRUM_ref}, allowing for the online merging of coincident events to be stored; the timestamp information is also saved and used in the offline analysis to perform an \textit{a posteriori} rejection of some possible random coincidences.

The data acquired in the first INDRA-FAZIA experiment have been analyzed in Ref~\cite{Ciampi2022} by studying the QP-QT isospin diffusion in the binary exit channel by means of the isospin transport ratio method \cite{Rami2000}, and they will be further inspected in the present paper, now with a focus on the QP breakup channel. In this campaign, the reactions for all of the four possible combinations of $^{64}$Ni and $^{58}$Ni have been studied for two different incident beam energies belonging to the Fermi regime, namely 32 and $52\,$MeV/nucleon; about $3\times10^7$ events have been acquired for each measured reaction. Such a complete set of reactions allows to study the role of the isospin gradient in the system by comparing the results of the symmetric and asymmetric reactions, also inspecting the possible consequences of the different dynamical features at the two energies.

\section{Data analysis}
For the following analysis, as done in Ref.~\cite{Ciampi2022}, we only consider the events with a total multiplicity $M\geq2$ of particles identified in FAZIA, thus excluding the elastic scattering, but also the events with only one ejectile in FAZIA and one (or more) ejectiles in INDRA.
We also perform the same preliminary selection, discarding the events violating charge and momentum conservation laws (corresponding to spurious coincidences) as well as strongly incomplete events, namely those with a total detected charge $Z_\text{tot}<12$. As a consequence of the selections, the following analysis is carried out on a subset of the total statistics including about 46\% (57\%) of all the events acquired for the reactions at $32\,(52)\,$MeV/nucleon. 
For all the systems, the majority of the events thus selected presents a multiplicity of medium-sized to heavy fragments (i.e., with $Z\geq5$) $M_{Z\geq5}=1$; such subclass represents around 70\% (65\%) of the events for the reactions at $32\,(52)\,$MeV/nucleon, corresponding to the binary exit channel studied in Ref.~\cite{Ciampi2022}. 

 \subsection{Selection of the QP breakup channel}
 \label{ssec:breakup_selection}
 In order to extend the analysis to the QP breakup channel, we hereby consider the $M_{Z\geq5}=2$ subclass, representing around 20\% (15\%) of the statistics at $32\,(52)\,$MeV/nucleon. The two $Z\geq5$ fragments are marked as heavy (HF) and light fragment (LF), depending on their atomic number, or on their mass number if $Z_\text{HF}=Z_\text{LF}$. Since we aim to carry out an analysis of isospin related observables on the two fission fragments, we also require them to be isotopically identified: as a consequence, since in this experiment INDRA could provide mass identification only up to $Z\approx4$, only the events with both the HF and the LF collected by FAZIA have been considered, representing about 40-50\% of the $M_{Z\geq5}=2$ subset.
 
\begin{figure}
  \centering
  \includegraphics[width=0.48\columnwidth]{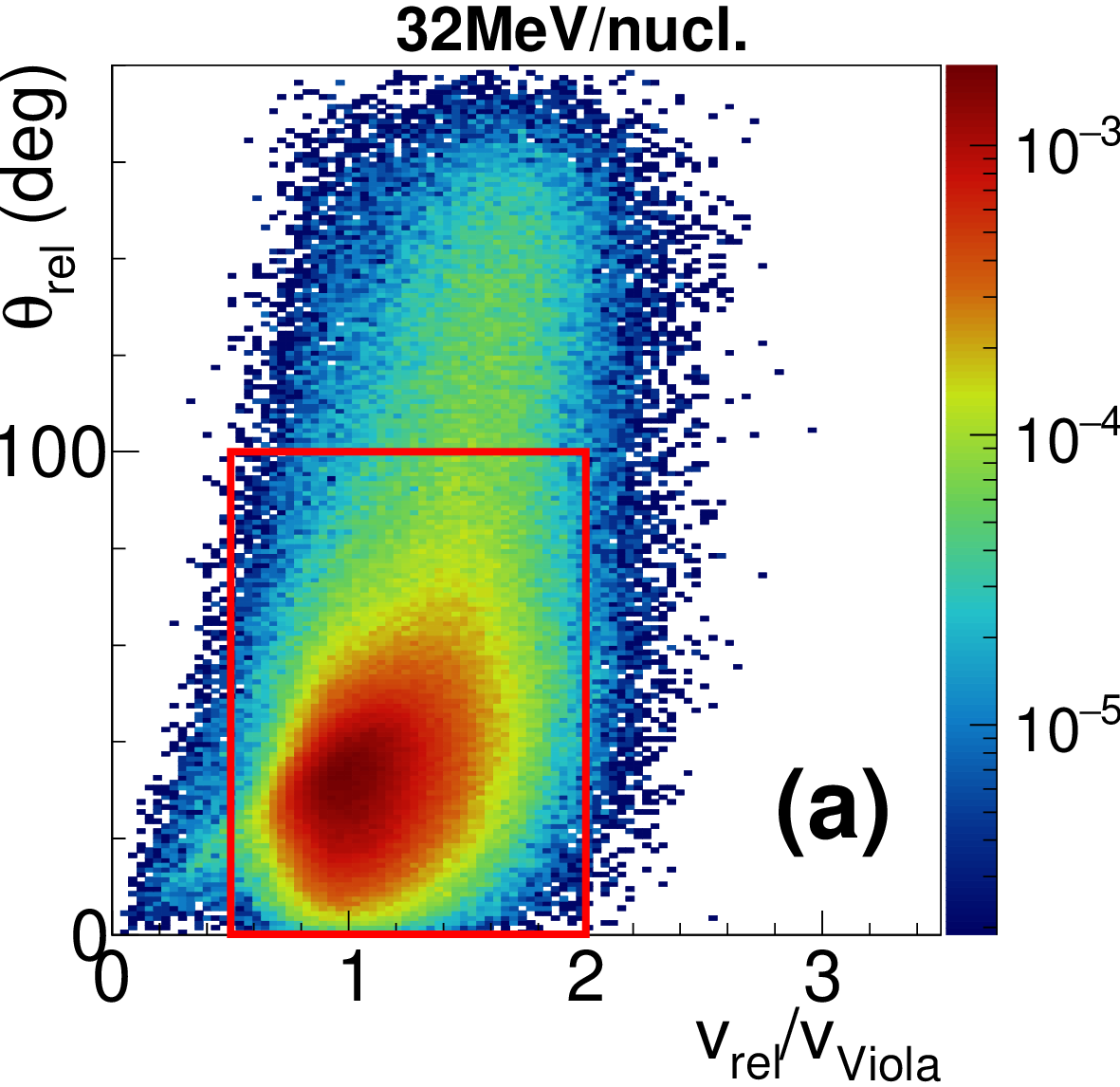}~~
  \includegraphics[width=0.48\columnwidth]{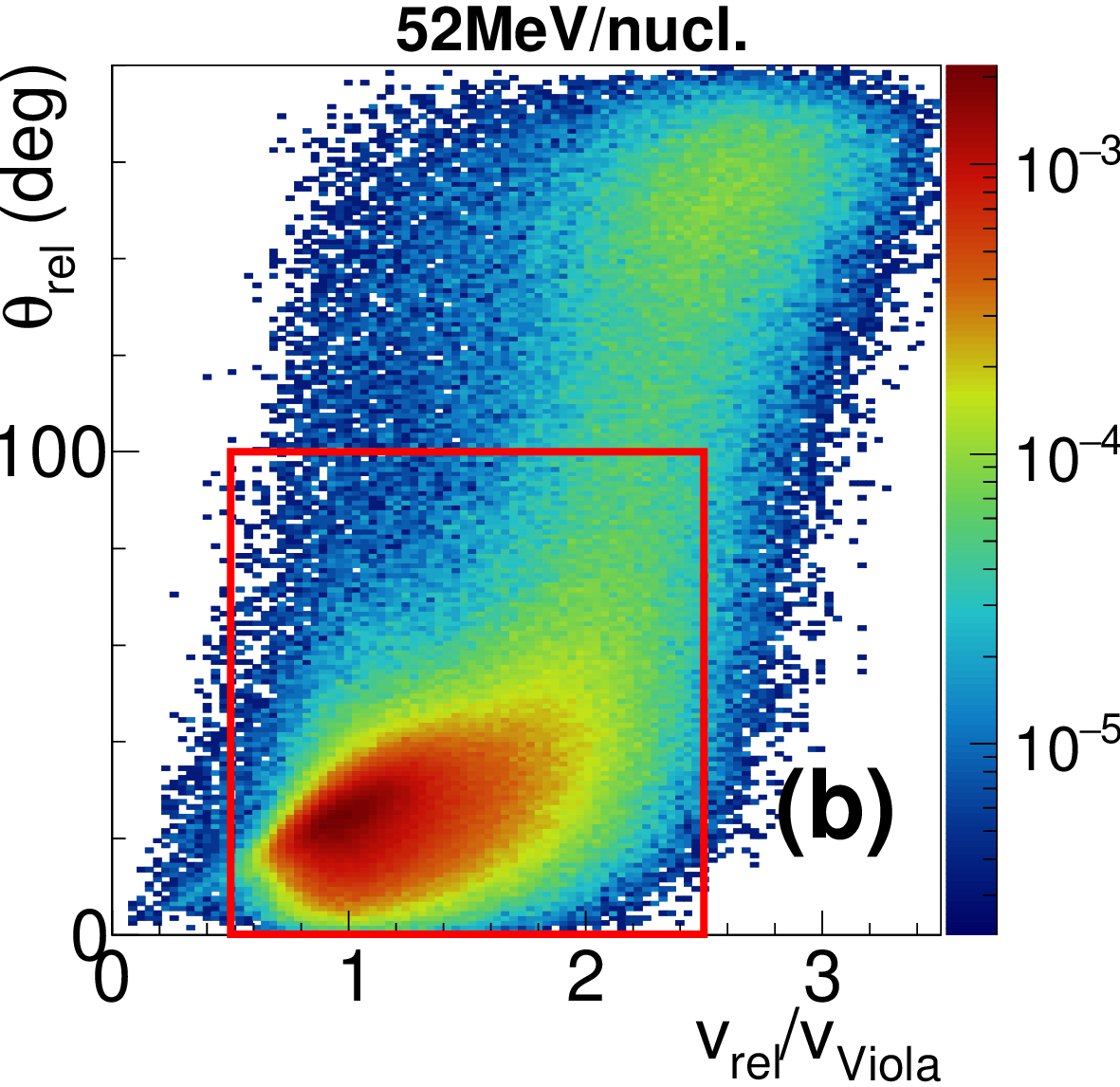} 
  \caption{Experimental $\theta_\text{rel}$ vs $v_\text{rel}/v_\text{Viola}$ correlations for the events with $M_{Z\geq5}=2$, for the reaction $^{58}$Ni+$^{58}$Ni at $32\,$MeV/nucleon (a) and at $52\,$MeV/nucleon (b). Similar results are obtained for the other three available systems at the respective beam energy. The plots are normalized to their integral.}
  \label{fig:thetarel_vrel}
 \end{figure}

 Among the events in such $M_{Z\geq5}=2$ subclass we select those compatible with a QP fission. In order to exclude ``spurious" events in which the QP remnant is detected together with a fragment of the QT, we exploit the correlation between the relative angle $\theta_\text{rel}$ between the velocities of the two fragments in the c.m. reference frame and the magnitude of their relative velocity $v_\text{rel}$ \cite{Piantelli2020}, divided by the relative velocity calculated according to the Viola systematics in the original prescription \cite{Viola1985}, that in Ref.~\cite{Piantelli2023} demonstrates to avoid distortions for the most asymmetric splits.
 Figure~\ref{fig:thetarel_vrel} shows, as an example, the correlations obtained for the $^{58}$Ni+$^{58}$Ni reaction at the two beam energies. %
 The events corresponding to the detection of such spurious QP-QT events are more likely to be located in the region at larger $\theta_\text{rel}$, closer to 180\textdegree, which is indeed more populated for the higher beam energy due to the fact that the beam boost facilitates the detection and identification of heavy fragments belonging to the QT phase space.
 Conversely, the region corresponding to small relative angles is more compatible with the sought after scenario of a QP fission event in which both the QP fission products have been detected. This observation is also supported by the fact that the typical relative velocities for $\theta_\text{rel}<100$\textdegree~are quite close to those calculated according to the fission systematics \cite{Viola1985}. We therefore select the events falling in the region delimited by the red line in Fig.\ref{fig:thetarel_vrel}, cutting out also those where the $v_\text{rel}$ between the two heavy fragments is too low to be compatible with their mutual Coulomb repulsion. 
 We have verified that the results presented henceforth are essentially stable against reasonable variations of these limits.
 
\begin{figure}
  \centering
  \includegraphics[width=0.48\columnwidth]{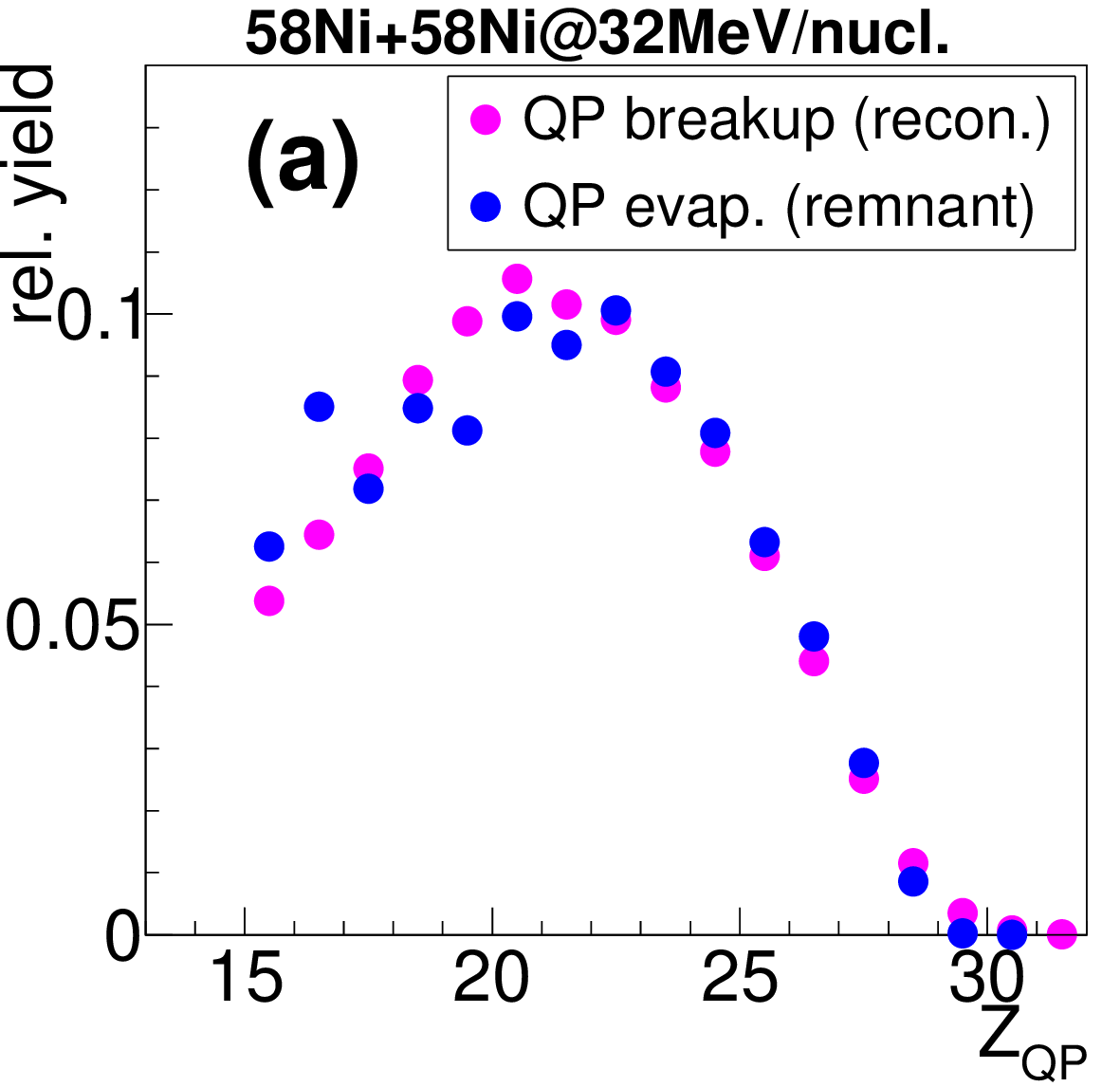}
  \includegraphics[width=0.48\columnwidth]{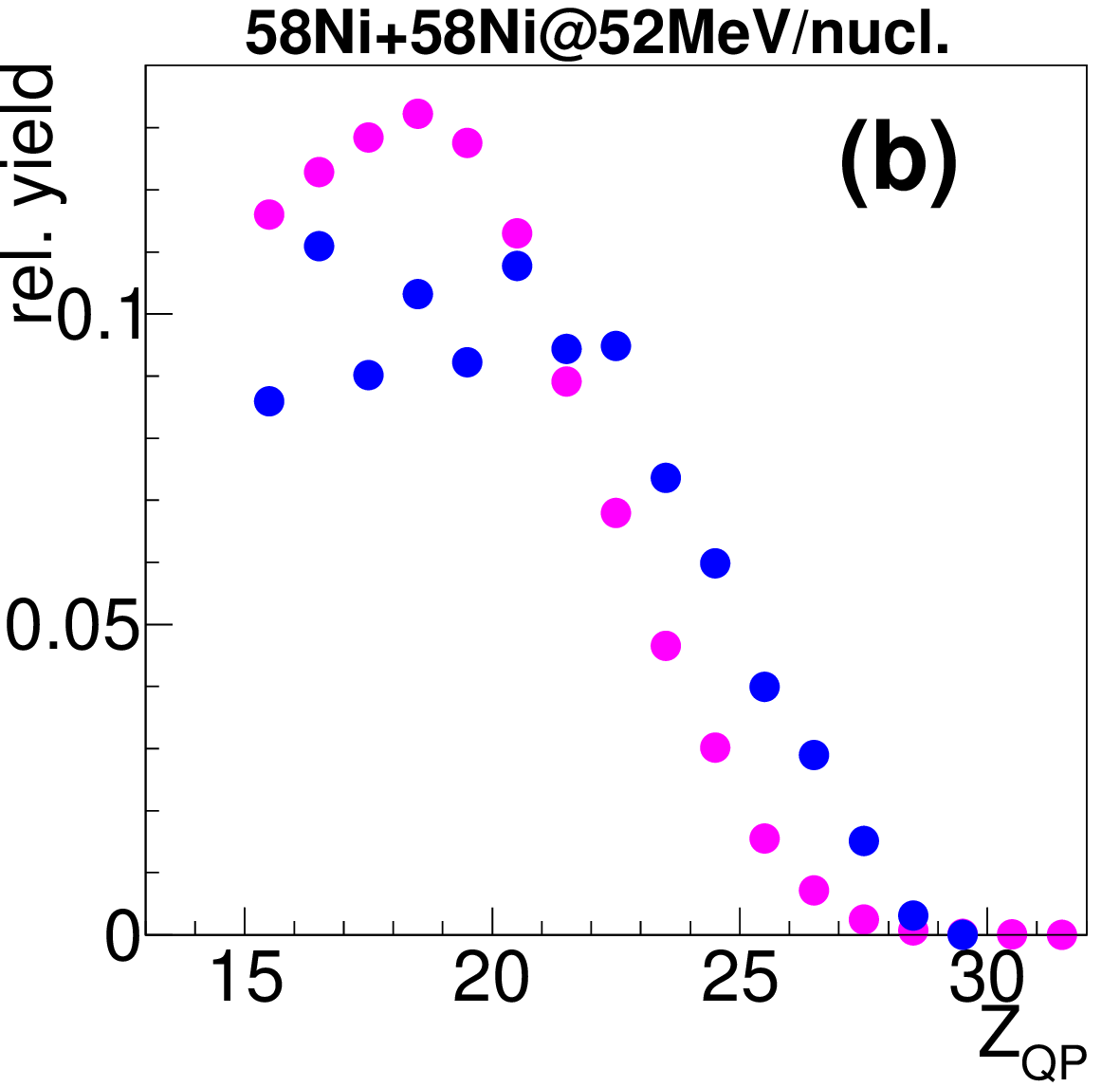}
  \includegraphics[width=0.48\columnwidth]{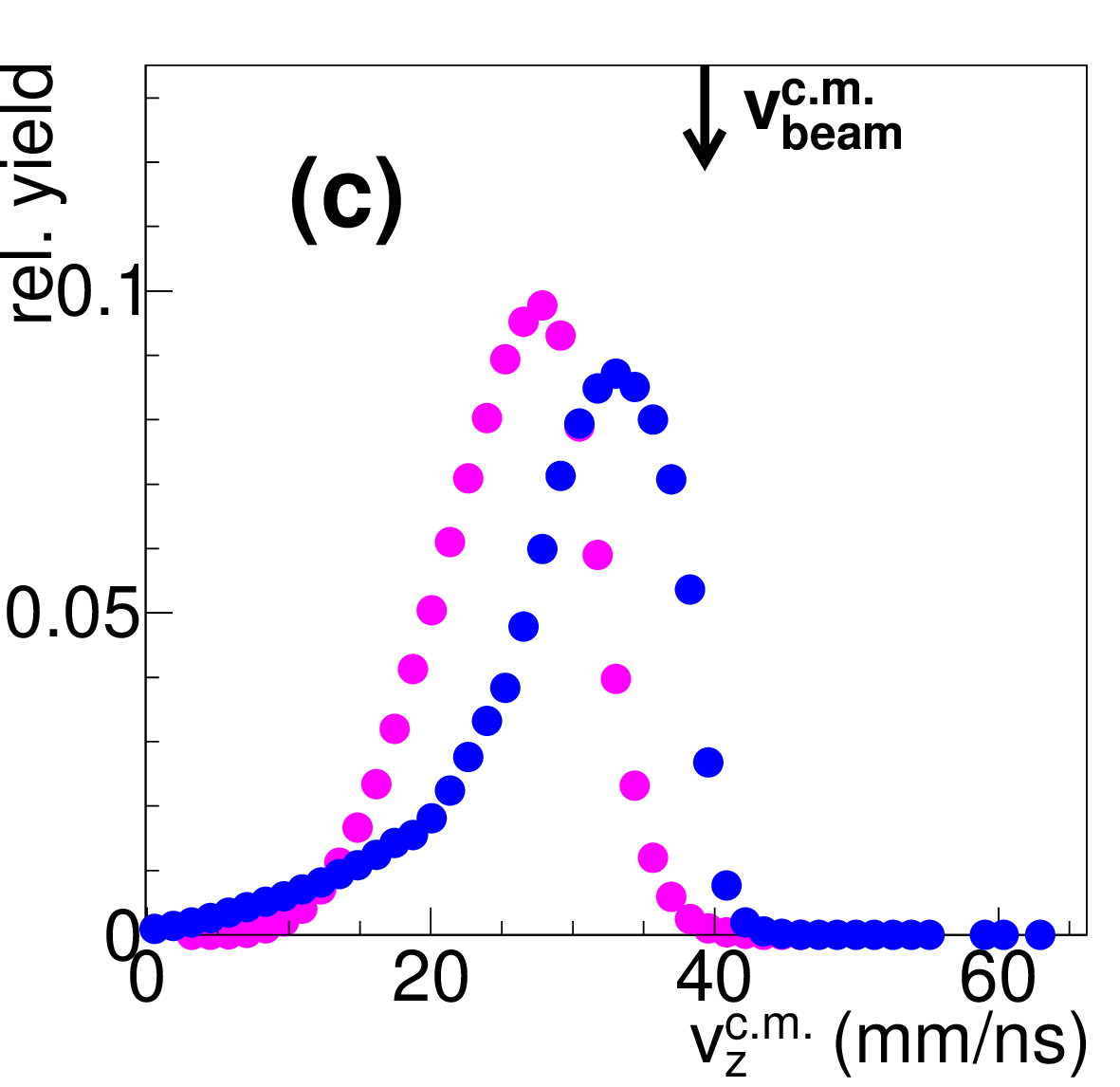}
  \includegraphics[width=0.48\columnwidth]{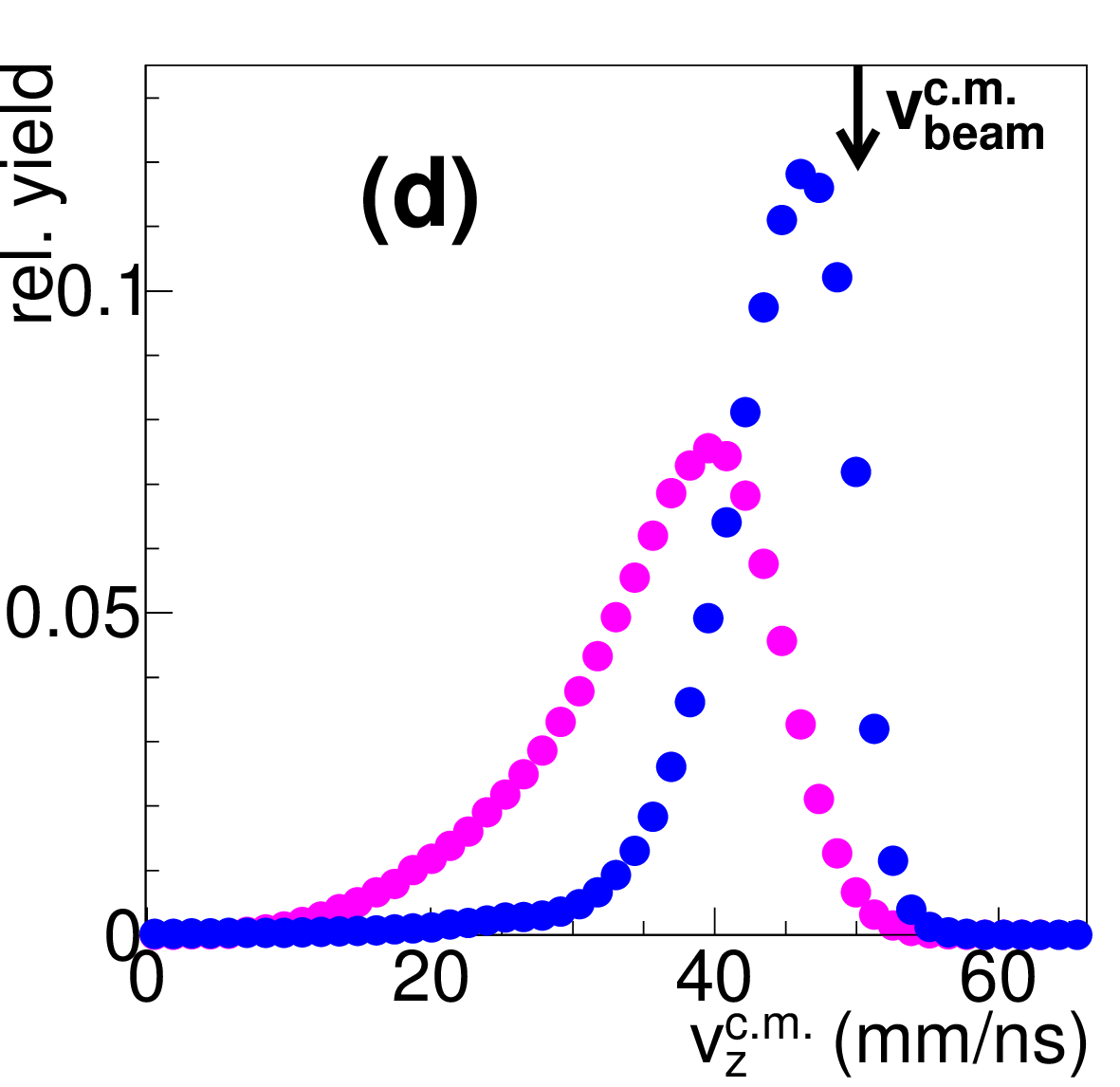}
  \caption{Comparison between the experimental charge distributions (top) and velocity distributions (bottom) obtained for the reconstructed QP in the breakup channel (magenta) and for the QP remnant in the binary channel (blue, same as in Ref.~\cite{Ciampi2022}), for the reaction $^{58}$Ni+$^{58}$Ni at $32\,$MeV/nucleon (left) and at $52\,$MeV/nucleon (right). The black arrows in plots (c),(d) indicate the original projectile velocity. The plots are normalized to their integral.}
  \label{fig:Z_vz_monodim}
 \end{figure}

 From the two daughter nuclei candidates, we ``reconstruct" the fissioning QP as a nucleus with charge number $Z_\text{rec}=Z_\text{HF}+Z_\text{LF}$, mass number $A_\text{rec}=A_\text{HF}+A_\text{LF}$, and velocity $\mathbf{v}$ equal to that of the c.m.~of the HF and the LF. The QP thus reconstructed is also required to be forward emitted in the reaction c.m.~reference frame ($v_{z}^\text{c.m.}>0$) and to have a charge number $Z_\text{rec}\geq15$. The latter condition is the same one imposed on the QP remnant in the evaporation channel of Ref.~\cite{Ciampi2022}: also in this case no substantial change in the final observations could be evidenced by varying this lower limit by $\pm3$ charge units. 
 In Fig.~\ref{fig:Z_vz_monodim} we show the charge [Figs.~\ref{fig:Z_vz_monodim}(a),(b)] and velocity [Figs.~\ref{fig:Z_vz_monodim}(c),(d)] distributions for the reconstructed QP (in magenta), for the $\sim5\times10^5$ events selected for the two reactions shown in Fig.~\ref{fig:thetarel_vrel}: the characteristics resemble those of a forward emitted heavy QP-like fragment (in the reaction c.m.~reference frame), with typical velocities tending to the original projectile one, indicated with a black arrow in the plots.
 The results for the reconstructed QP are also compared to what obtained for the QP remnant in the binary exit channel (in blue, taken from Ref.~\cite{Ciampi2022}).
 The charge distributions in the two cases are quite similar, while a slightly more evident shift is present between the velocity distributions in the two channels: for the QP evaporation channel, less dissipative events are selected (see, e.g., the differences between the velocity distributions), including more peripheral collisions. A more extensive comparison between the two channels will be presented in Sec.~\ref{ssec:QPr-QPb_comparison}. However, we stress that the similarities between the reconstructed QP and the typical QP remnants support the fission event selection procedure.
 
\begin{figure}
  \centering
  \includegraphics[width=0.8\columnwidth]{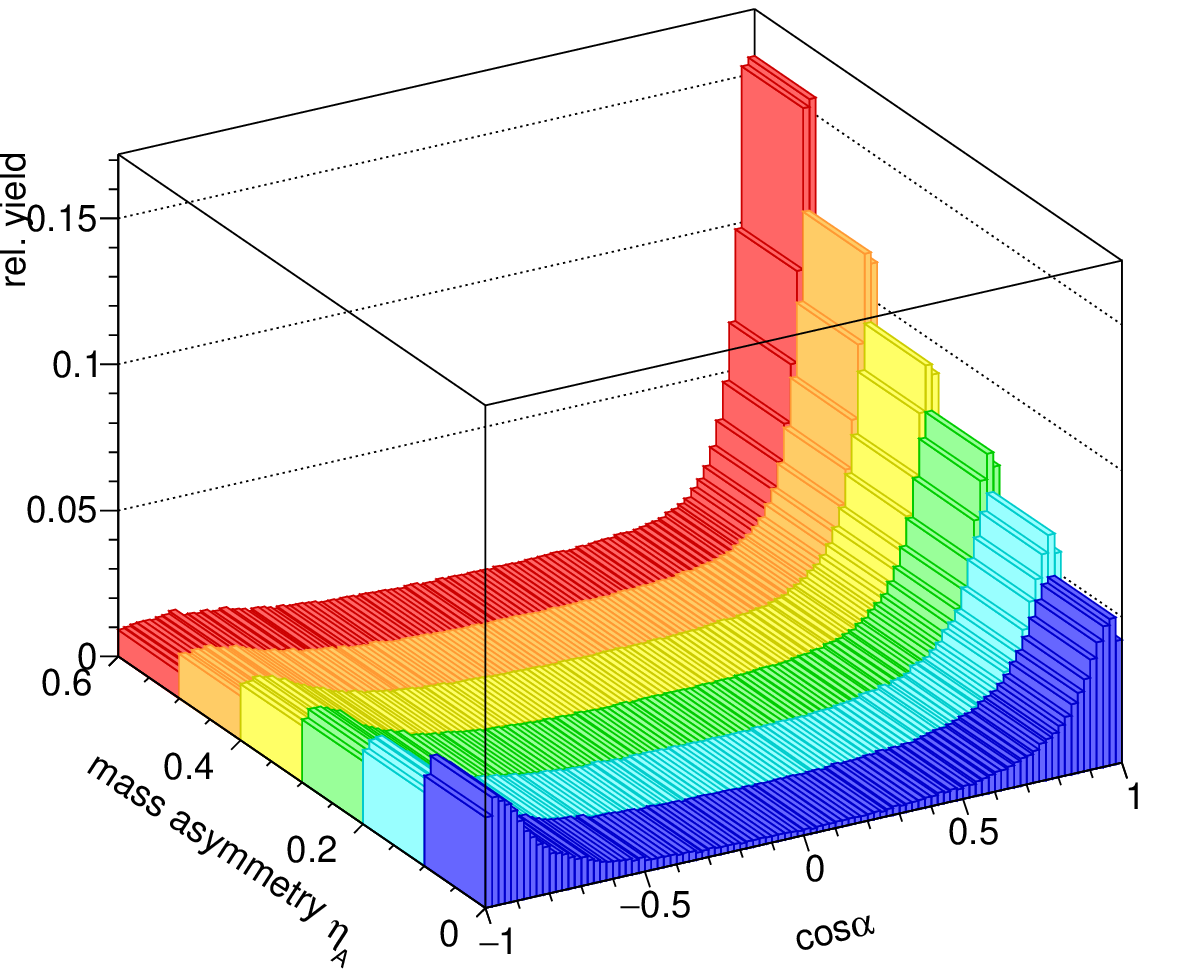}
  \caption{Experimental $\cos\alpha$ distributions obtained for the reaction $^{58}$Ni+$^{58}$Ni at $32\,$MeV/nucleon for six intervals of mass asymmetry of the split. Similar results are obtained for the other systems. The histogram for each $\eta_A$ bin is separately normalized to its integral.}
  \label{fig:alpha_eta}
 \end{figure}

\begin{figure*}
     \centering
     \includegraphics[width=0.23\textwidth]{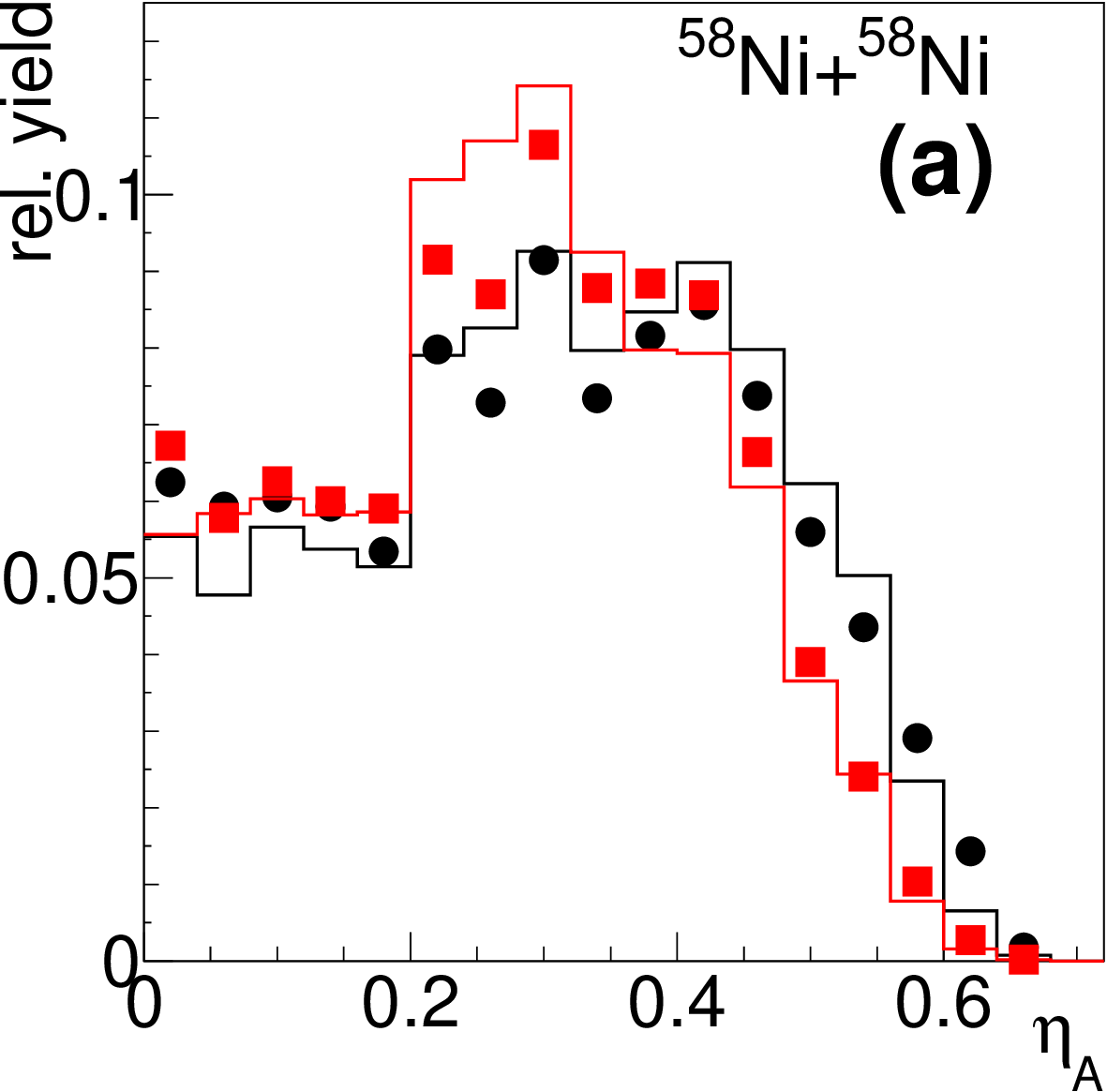}~~
     \includegraphics[width=0.23\textwidth]{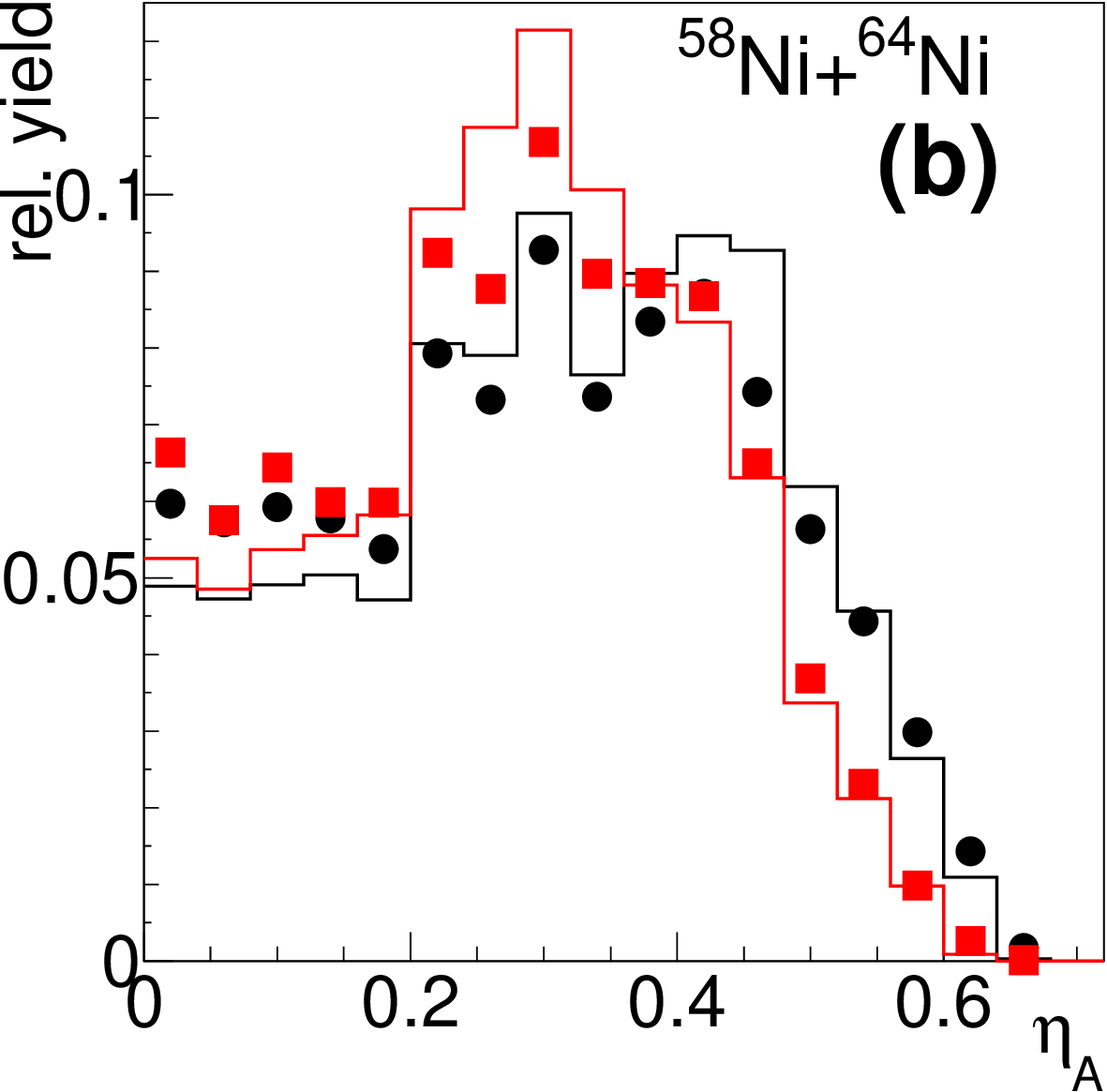}~~
     \includegraphics[width=0.23\textwidth]{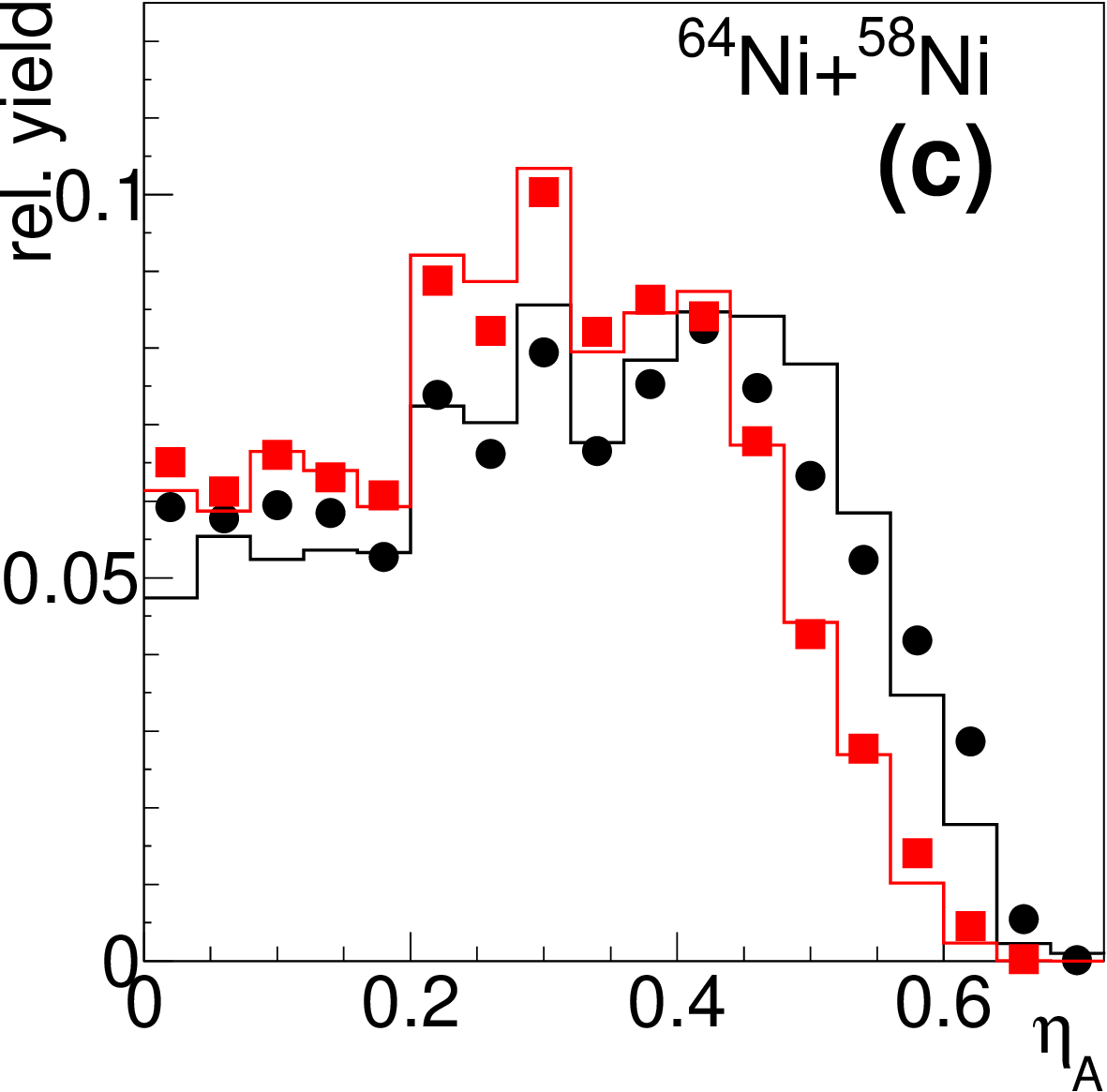}~~
     \includegraphics[width=0.23\textwidth]{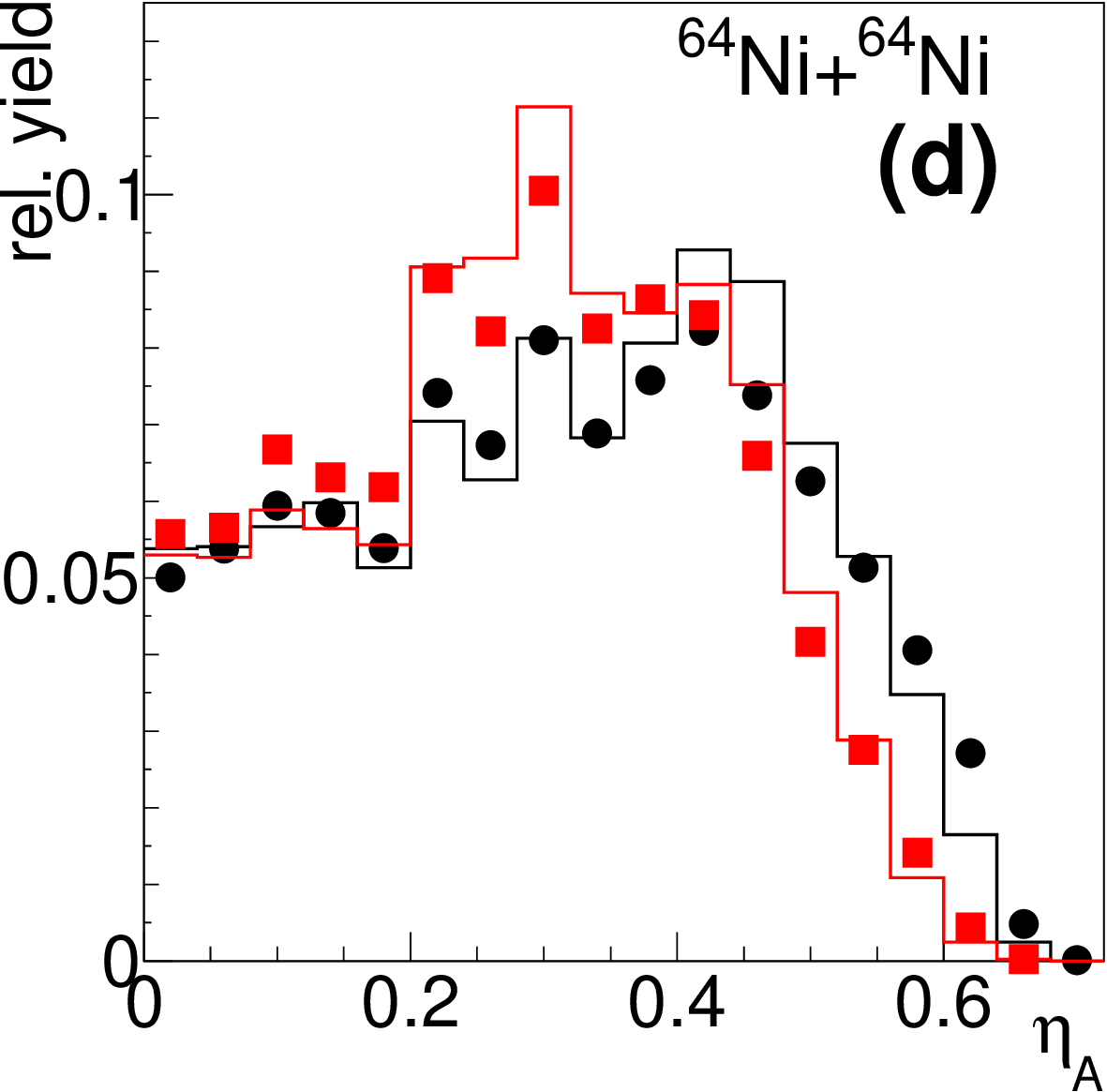}\\~\\
     \includegraphics[width=0.23\textwidth]{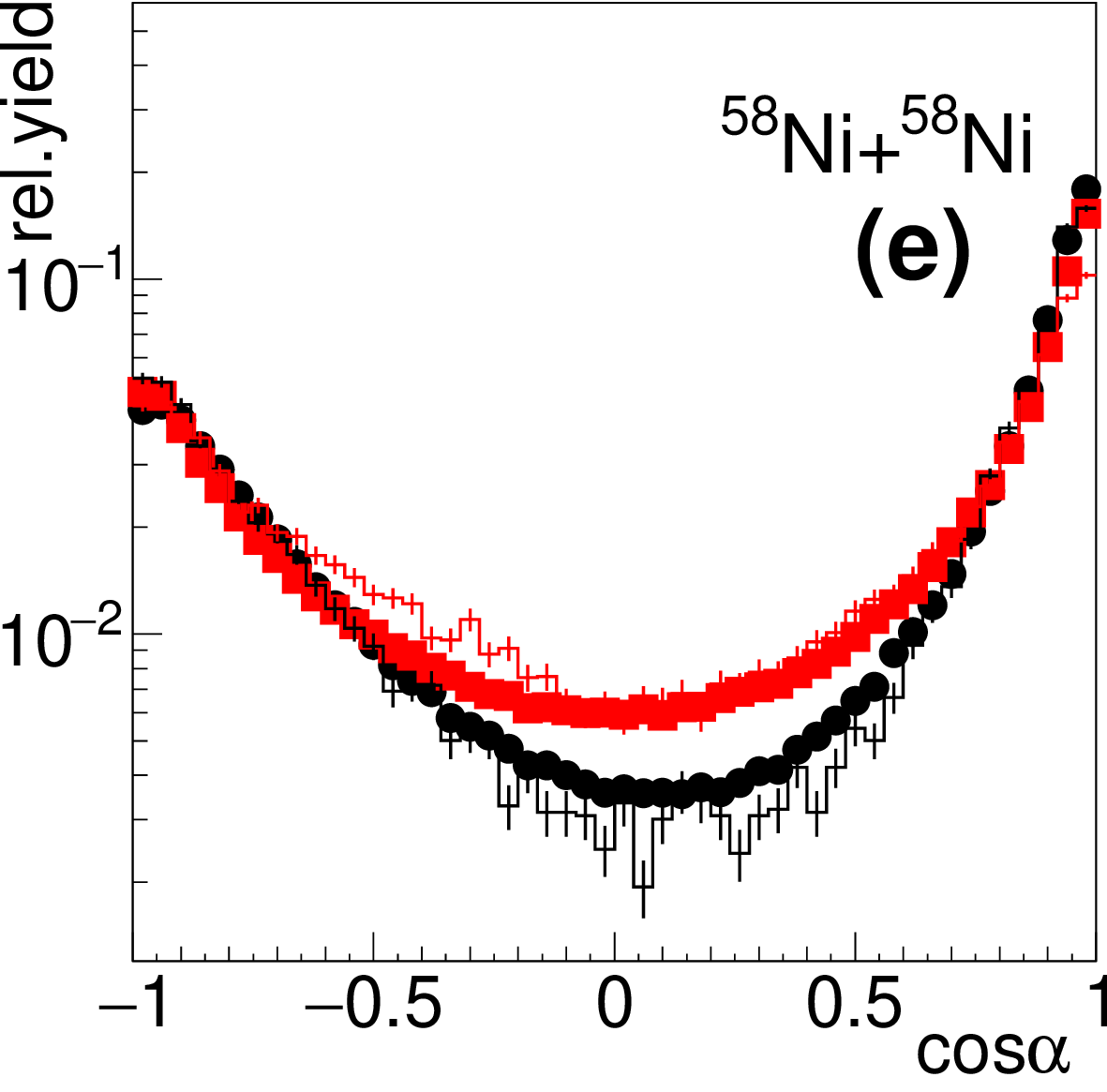}~~
     \includegraphics[width=0.23\textwidth]{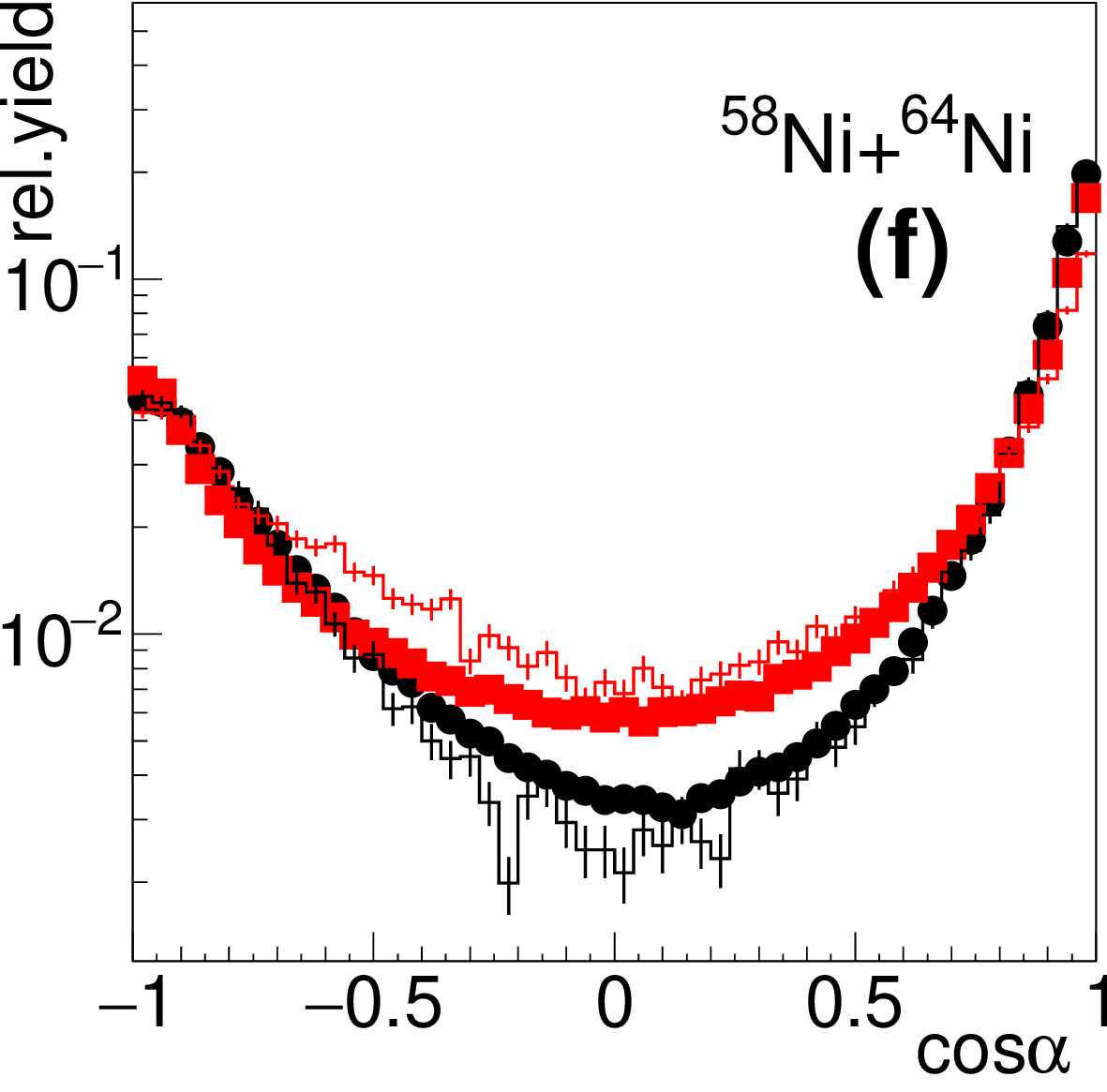}~~
     \includegraphics[width=0.23\textwidth]{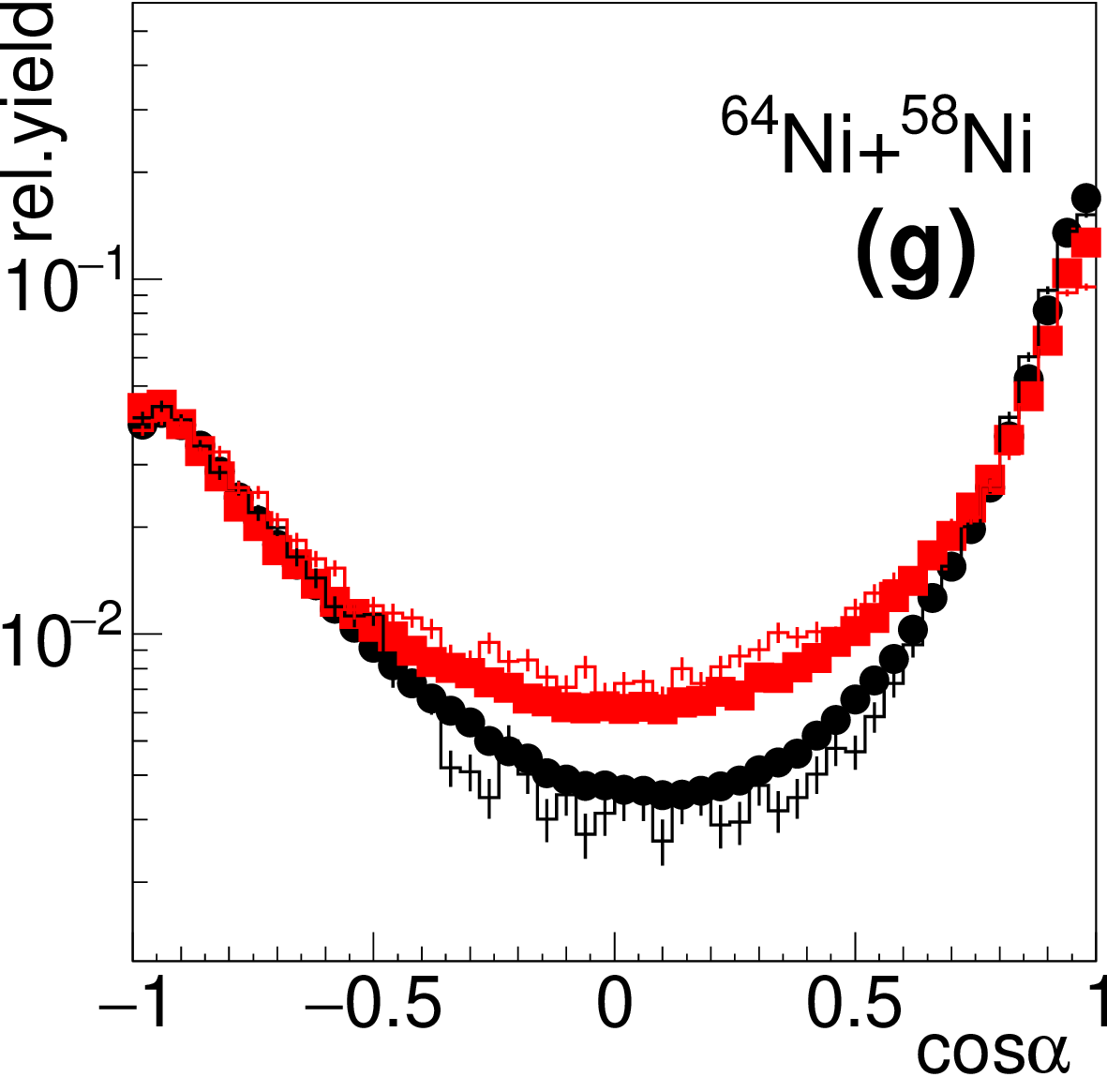}~~
     \includegraphics[width=0.23\textwidth]{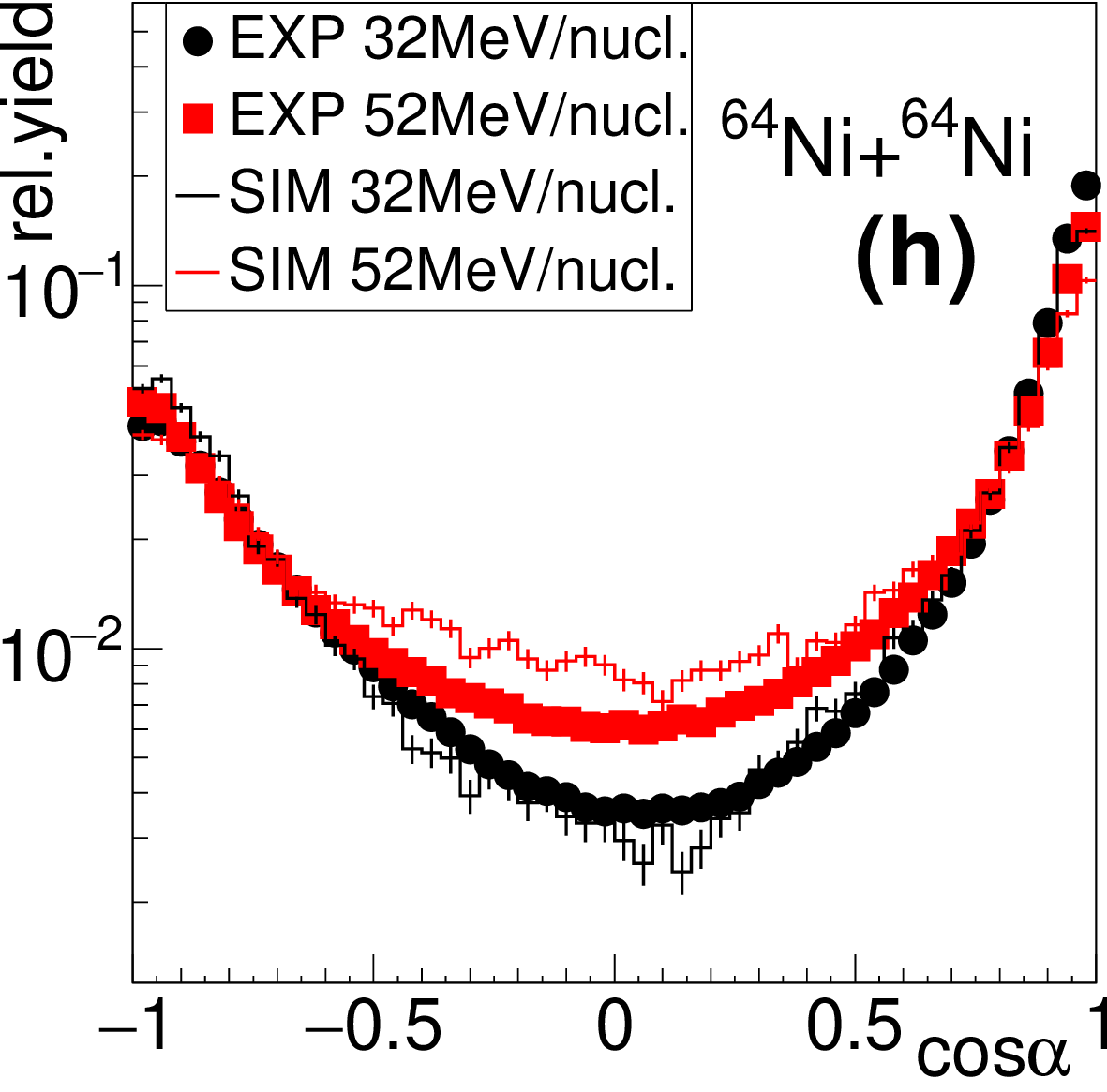}
     \caption{Experimental and simulated distributions of (a)-(d) the mass asymmetry $\eta_A$ and (e)-(h) of the cosine of the $\alpha$ angle obtained for all the studied reactions: [(a),~(e)] $^{58}$Ni+$^{58}$Ni, [(b),~(f)] $^{58}$Ni+$^{64}$Ni, [(c),~(g)] $^{64}$Ni+$^{58}$Ni and [(d),~(h)] $^{64}$Ni+$^{64}$Ni. The reactions at $32\,$MeV/nucleon ($52\,$MeV/nucleon) are plotted in black (red). The histograms are normalized to their integral. Experimental data are plotted with solid markers, while AMD+\textsc{Gemini++} predictions are drawn with a line. The same legend applies to all plots.}
     \label{fig:eta_alpha_modelpredic}
 \end{figure*}

 This selection of events can be generally interpreted as QP fissions, independently of the nature of the split. As anticipated, according to the available observations in the literature \cite{Glassel1983,Stefanini1995,Casini1993,Bocage2000,DeFilippo2005,Piantelli2020}, the favoured emission of the LF towards the reaction c.m.~associated to more asymmetric splits is a characteristic feature of the dynamical fission.
 We therefore inspect the mass asymmetry of the two fission fragments $\eta_A= \frac{A_\text{HF}-A_\text{LF}}{A_\text{HF}+A_\text{LF}}$, in relation to the orientation of the HF-LF split, expressed by means of the aforementioned $\alpha$ angle, extracted as the angle between the direction of the QP velocity in the c.m.~reference frame $\mathbf{v}$ and the direction of the relative velocity between the HF and the LF $\mathbf{v}_\text{rel} = \mathbf{v}_\text{HF}-\mathbf{v}_\text{LF}$:
 \begin{equation}
  \label{eq:alpha_definition}
  \alpha=\arccos\biggl(\frac{\mathbf{v}\cdot\mathbf{v}_\text{rel}}{|\mathbf{v}|\cdot|\mathbf{v}_\text{rel}|}\biggr)
 \end{equation}
 Note that the configuration in which the LF is backward emitted, perfectly aligned with the QP-QT axis, corresponds to $\alpha=0$\textdegree.
 The experimental distributions of $\cos\alpha$ for six $\eta_A$ intervals are shown in Fig.~\ref{fig:alpha_eta} for one of the measured reactions. 
 Similar results are obtained for the other systems. For the most symmetric mass splits with $\eta_A\sim0$, we find a rather symmetric $\cos\alpha$ distribution. On the contrary, increasingly asymmetric breakups tend to occur in HF-LF configurations preferentially aligned with the QP-QT axis, with the LF emitted backward, as indicated by the evident peak at $\cos\alpha\sim1$. Such a characteristic feature indicates a predominant contribution of dynamical fission processes on the total fission statistics.
 The experimental mass asymmetry distributions for all the measured reactions are plotted with solid markers in Figs.~\ref{fig:eta_alpha_modelpredic}(a)-(d). The corresponding experimental $\cos\alpha$ distributions are shown in Figs.~\ref{fig:eta_alpha_modelpredic}(e)-(h), solid markers. The mass splits tend to be quite asymmetric, with $\eta_A$ values preferentially greater than 0.2, even if the statistics for $\eta_A>0.5$ rapidly decreases: however, we want to stress that the choice of the threshold $Z>4$ for the LF and, to a lesser extent, the mass identification limits for the HF put an upper bound to the accessible $\eta_A$, also considering the relatively small projectile size. Probably also due to the condition imposed on the minimum value of $Z_\text{LF}$, the mass splits for the reactions at 52~MeV/nucleon, which produce lighter reconstructed QPs (see Fig.~\ref{fig:Z_vz_monodim}), are less asymmetric than those at the lower beam energy.
 
 The set of reactions studied in this experiment have been also simulated by means of the antisymmetrized molecular dynamics (AMD) \cite{Ono1992} model coupled to \textsc{Gemini++} \cite{Charity2010} as afterburner. About 20000 primary events have been produced with the version of AMD described in Ref.~\cite{Piantelli2019}, assuming an asy-stiff parametrization of the symmetry energy term of the NEoS\footnote{For the results presented in this paper, the asy-soft predictions are comparable to the asy-stiff ones, and are thus omitted for brevity.} and a triangular distribution of the impact parameter up to $11.6\,$fm, slightly larger than the grazing impact parameter $b_\text{gr}$ for all reactions. The AMD simulations have been run until $500\,$fm/$c$ ($1\,\text{zs}\approx300\,\text{fm}/c$), then, for each primary event, 100 secondary events have been generated with \textsc{Gemini++} to increase the statistics. The simulated events thus produced have been then filtered according to a realistic software replica of the experimental apparatus and analyzed by means of the same code and same selections used for the experimental data. 
 The corresponding simulated $\eta_A$ and $\cos\alpha$ distributions are presented on top of the experimental ones in the respective plots of Fig.~\ref{fig:eta_alpha_modelpredic} (black and red lines). 
 The main fission features are satisfactorily reproduced by the simulations at both energies. Since the model calculation demonstrates to be able to reasonably reproduce the experimental data also for the QP reconstructed from the breakup fragments (as for the QP remnant, as shown in Ref.~\cite{Ciampi2022}), we exploit the AMD+\textsc{Gemini++} simulations as an additional tool to investigate the origin of the QP fissions. Within the model predictions, we can easily distinguish between the fissions produced by AMD, whose origin is therefore dynamical, and those produced by \textsc{Gemini++} according to a statistical approach. In our filtered simulations, about 90\% (85\%) of the fission events are labelled as dynamical for the reactions at $32\,$MeV/nucleon ($52\,$MeV/nucleon). 
 
 In summary, according to the evaluations done both on experimental and simulated data, we can safely assume that a large majority of the selected fission events have a dynamical origin, in agreement with other findings \cite{Stefanini1995,Bocage2000,DeFilippo2005,Piantelli2020,RodriguezManso2017}.

 \subsection{Isospin analysis}
 \label{ssec:isospin_analysis}
  Taking advantage of the isospin information provided by the apparatus, in close similarity with the QP evaporation channel analysis of Ref.~\cite{Ciampi2022} for the same dataset, we exploit the average neutron to proton ratio $\langle N/Z\rangle$ of the QP reconstructed from its two fission fragments as isospin related observable. As order parameter, we employ the reduced reconstructed QP momentum $p_\text{red}$, defined as the ratio between the component along the beam axis of the momentum $p_z^\text{QP}$ of the reconstructed QP and the original projectile momentum $p_\text{beam}$, both of them in the reaction c.m.~reference frame: this quantity is completely analogous to the $p_\text{red}$ defined for the QP remnant in Refs.~\cite{Camaiani2021,Ciampi2022}, where its correlation with the reduced impact parameter $b_\text{red}=b/b_\text{gr}$ has been demonstrated.
  We recall that such correlation has been proven independent of the studied system, and only slightly dependent on the bombarding energy, according to AMD+\textsc{Gemini++} simulations. As in Ref.~\cite{Ciampi2022}, we point out that also here we limit our evaluations down to $p_\text{red}\sim0.4$, that we identified as the lowest value for which the $p_\text{red}$ vs $b_\text{red}$ correlation can be considered reliable, according to AMD+\textsc{Gemini++}. 
  
  \begin{figure}
   \centering
   \includegraphics[width=0.8\columnwidth]{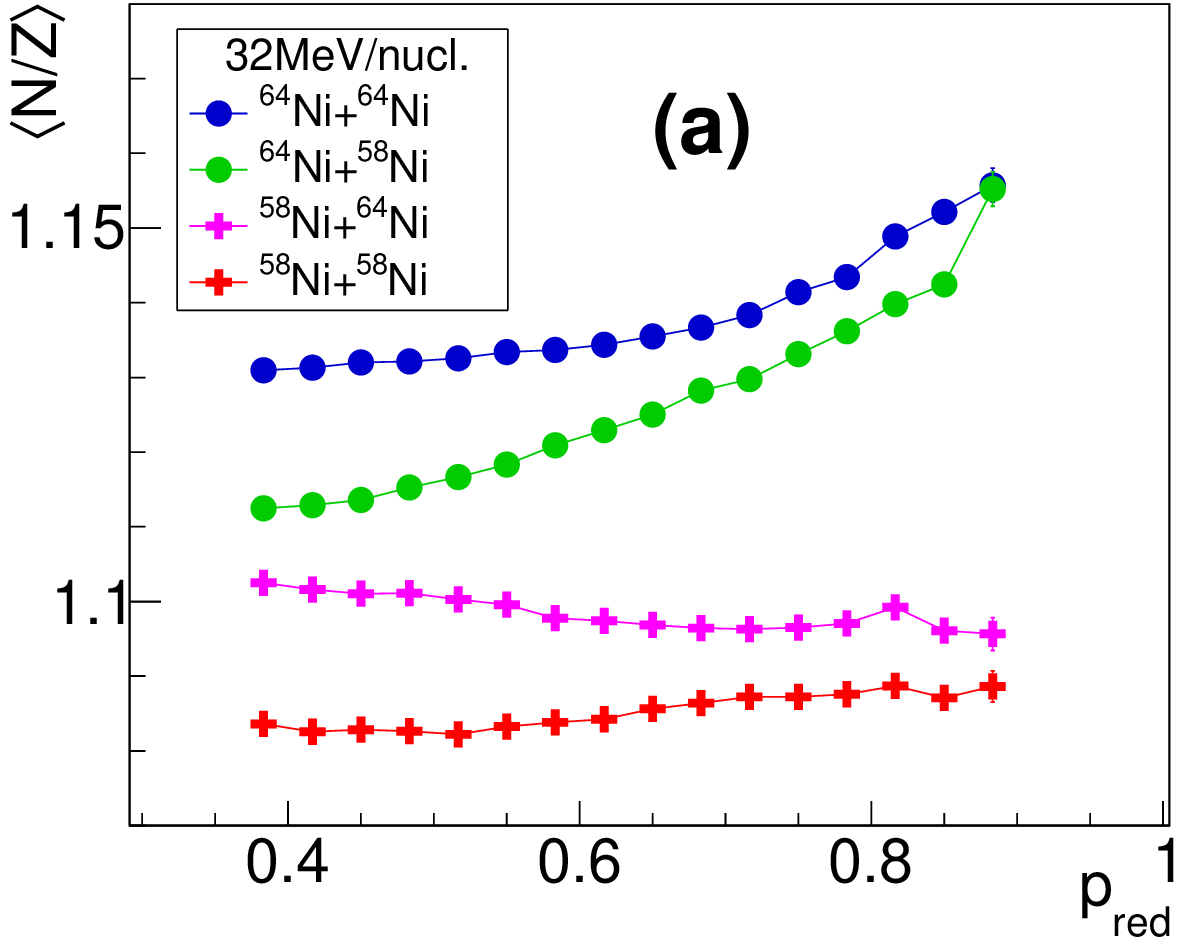}\\~\\
   \includegraphics[width=0.8\columnwidth]{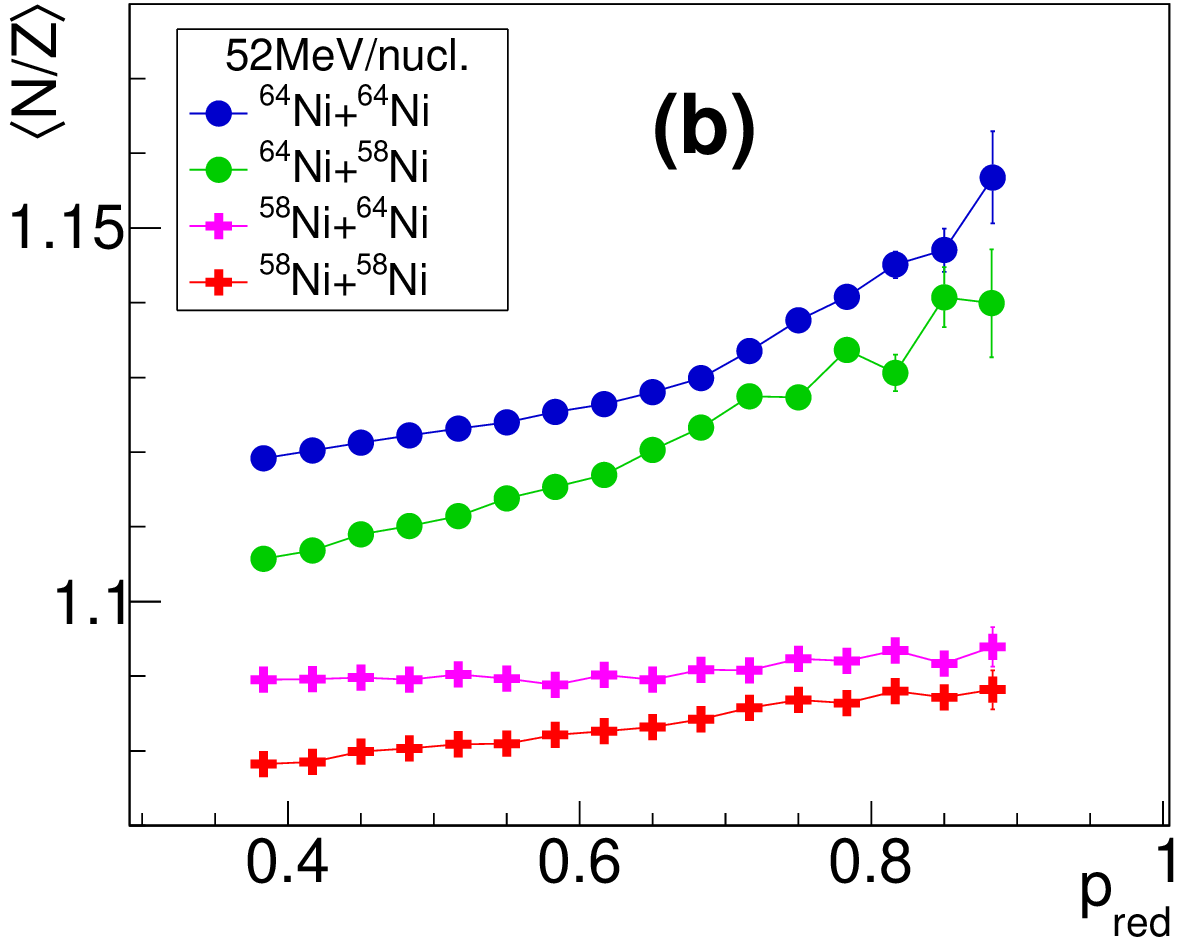} 
   \caption{QP breakup channel: experimental neutron to proton ratio $\langle N/Z\rangle$ of the reconstructed QP reported as a function of $p_\text{red}$ for the four reactions at $32\,$MeV/nucleon (a) and at $52\,$MeV/nucleon (b). Only statistical errors are shown.}
   \label{fig:NsuZ_pred_QPb}
  \end{figure}
  In Fig.~\ref{fig:NsuZ_pred_QPb} the experimental average neutron to proton ratios $\langle N/Z\rangle$ of the reconstructed QP as a function of $p_\text{red}$ are shown for the reactions at $32\,$MeV/nucleon [Fig.~\ref{fig:NsuZ_pred_QPb}(a)] and at  $52\,$MeV/nucleon [Fig.~\ref{fig:NsuZ_pred_QPb}(b)].
  We observe a clear hierarchy of the results for the four systems, starting from the most neutron rich symmetric reaction, down to the less neutron rich symmetric one. 
  Moreover, we notice a characteristic behavior of the two asymmetric (mixed) systems: in fact, in the most peripheral collisions (i.e., for larger $p_\text{red}$ values) the $\langle N/Z\rangle$ of the reconstructed QP tends to assume a value quite close to that obtained for the corresponding symmetric reaction induced by the same projectile (indicated by the same marker shape in Fig.~\ref{fig:NsuZ_pred_QPb}), while for increasing centrality, a difference gradually arises between them. As a consequence, the plots for the two asymmetric systems tend to a similar value at the lowest $p_\text{red}$ intervals here inspected. This evidence can be ascribed to the trend towards isospin equilibrium between projectile and target in the asymmetric reactions due to the isospin diffusion mechanism during the interaction.
  
\begin{figure}
   \centering
   \includegraphics[width=0.8\columnwidth]{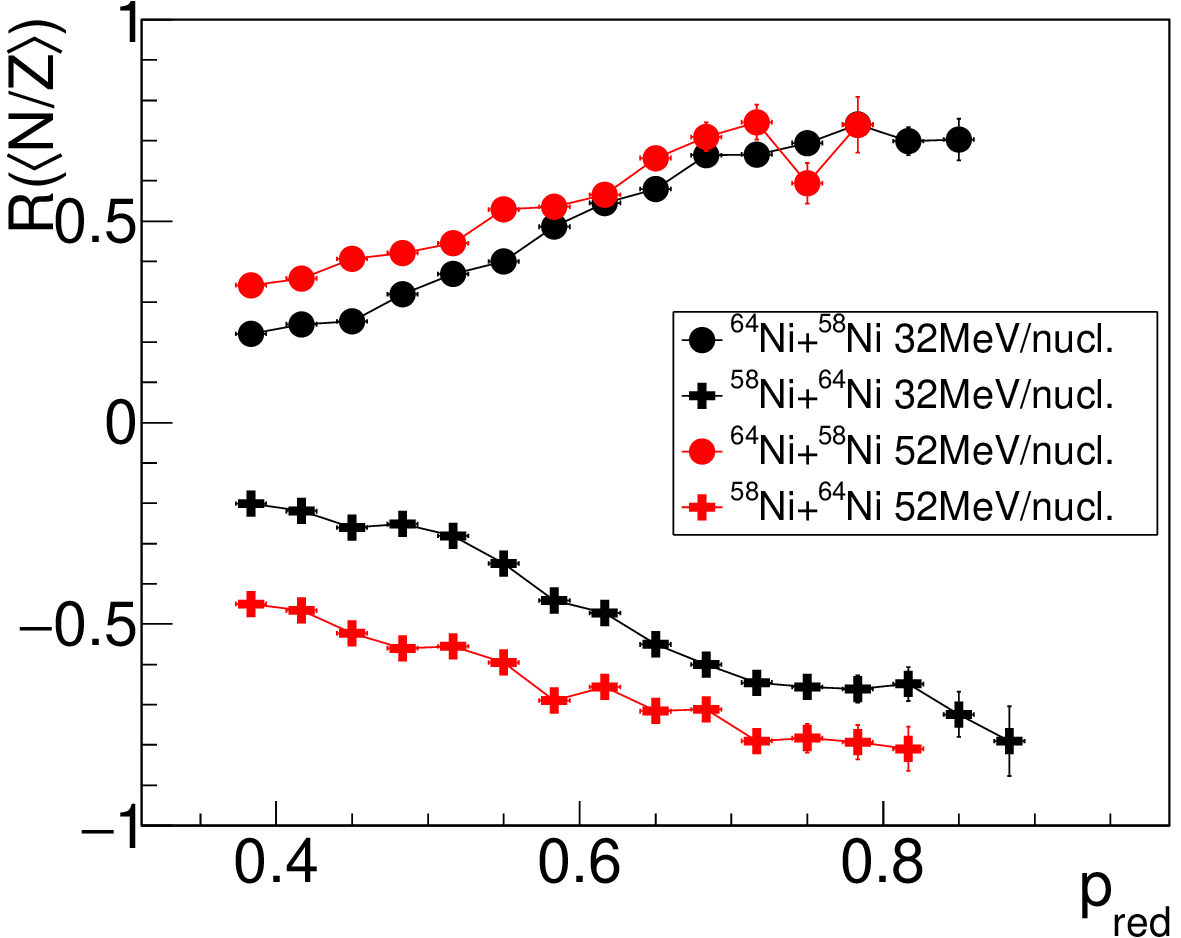}
   \caption{QP breakup channel: isospin transport ratio calculated with the $\langle N/Z\rangle$ of the reconstructed QP reported as a function of $p_\text{red}$ (experimental data). The results for the asymmetric reactions $^{64}$Ni+$^{58}$Ni and $^{58}$Ni+$^{64}$Ni at $32\,$MeV/nucleon ($52\,$MeV/nucleon) are plotted as black (red) symbols. Only statistical errors are shown.}
   \label{fig:imbratio_QPb_pred}
  \end{figure}

  We can highlight this signature of isospin diffusion by exploiting the isospin transport ratio method \cite{Rami2000}, widely used and studied in the literature \cite{May2018,Tsang2004,Tsang2009,Liu2007,Sun2010,Napolitani2010,Camaiani2020,Camaiani2021,Mallik2022,Fable2023} as a tool to study the isospin equilibration phenomenon, possibly enhancing the sensitivity to different assumptions on the $E_{sym}$ functional. With specific reference to the systems investigated in our experiment, the isospin transport ratio is defined as:
  \begin{equation}
   \label{eq:imbratio}
   R(X_{i}) = \frac{2X_i-X_{6464}-X_{5858}}{X_{6464}-X_{5858}}
  \end{equation}
  where $i$ can be one of the four reactions and $X$ is an isospin sensitive observable: in the following we will use $X\equiv\langle N/Z\rangle$ of the reconstructed QP. The values $R(X_{i})=\pm1$ obtained by construction for the two symmetric systems represent the absence of isospin diffusion. On the other hand, the two ``branches" of the ratio obtained for the two asymmetric reactions show the evolution of the isospin observable $X$ with respect to the two symmetric systems, which are used as a reference. The condition of complete isospin equilibration would be indicated by the result $R(X_{6458})=R(X_{5864})$.
  Figure~\ref{fig:imbratio_QPb_pred} shows the experimental $R_i(\langle N/Z\rangle)$ for the QP breakup channel as a function of $p_\text{red}$, for both asymmetric reactions (plotted with different markers associated to the different projectiles as in Fig.~\ref{fig:NsuZ_pred_QPb}) and for both beam energies ($32\,$MeV/nucleon in black, and $52\,$MeV/nucleon in red).
  Here, as already emerging from Fig.~\ref{fig:NsuZ_pred_QPb}, we can observe a clear evolution towards isospin equilibrium with increasing centrality, with the two branches starting from $R_i(\langle N/Z\rangle)$ values close to $\pm1$ for the most peripheral reactions, and tending towards each other with decreasing $p_\text{red}$. In the range of centrality explored in this paper, the full equilibration $R(X_{6458})=R(X_{5864})$ is not reached.
  On the other side, we also notice that $|R_i(\langle N/Z\rangle)|=1$ is not reached as well: however, we have checked that, according to the model predictions and to experimental results, the QP breakup channel is well populated only for less peripheral collisions, and the statistics drops as $b_\text{red}$ approaches 1. 
  Since this characteristics has been observed both in the filtered AMD+\textsc{Gemini++} simulations and in the unfiltered model predictions for the whole $4\pi$ solid angle, it does not seem to be related to the apparatus acceptance, but rather to the properties of this exit channel. Besides, it is quite feasible that the breakup channel may be associated to relatively large energy dissipated in the system: indeed, since in this mass region the $Q$-value for a fission process is significantly negative, an energy well above the activation energy for fissioning is needed in order to promptly access this channel.

  \begin{figure}
   \centering
   \includegraphics[width=0.8\columnwidth]{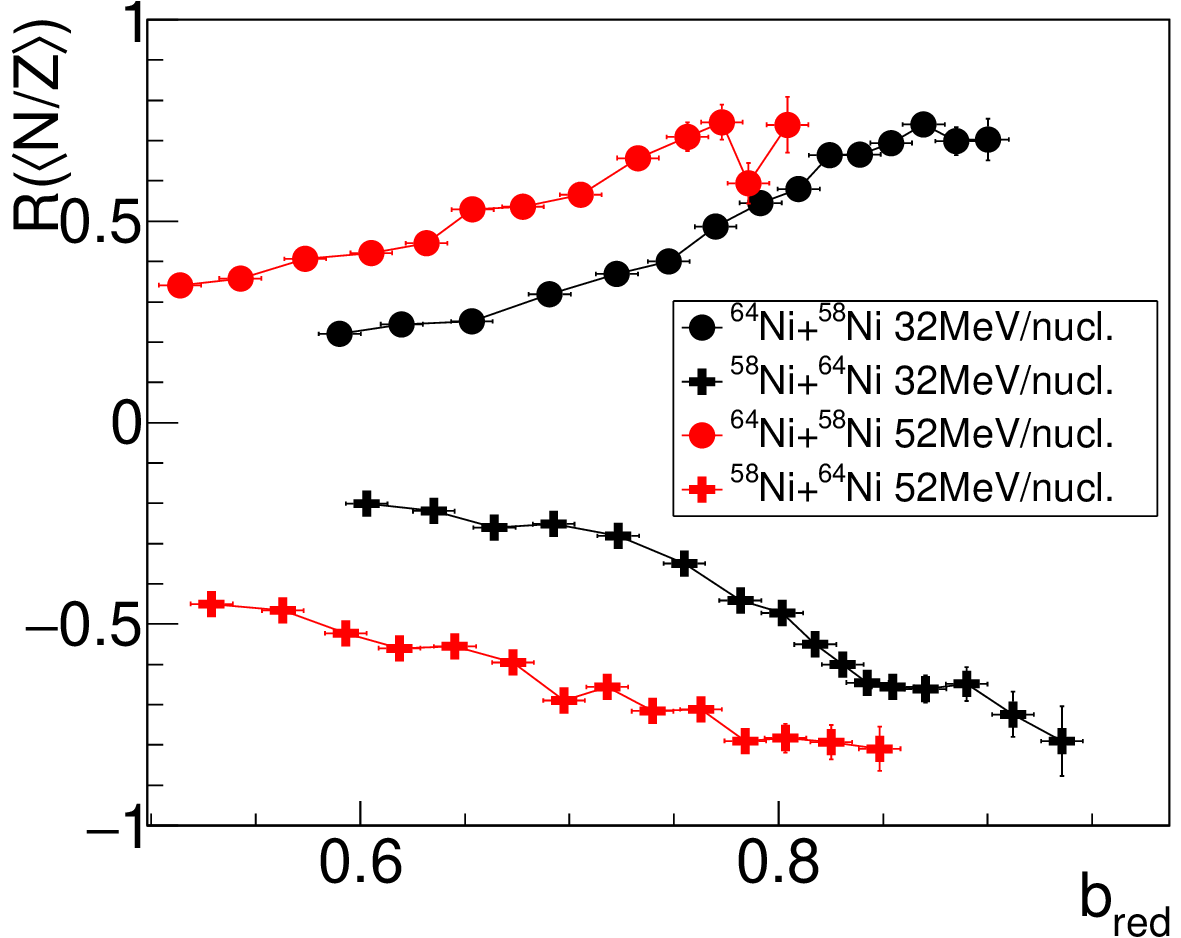}
   \caption{QP breakup channel: isospin transport ratio calculated with the $\langle N/Z\rangle$ of the reconstructed QP reported as a function of $b_\text{red}$. The results for the asymmetric reactions $^{64}$Ni+$^{58}$Ni and $^{58}$Ni+$^{64}$Ni at $32\,$MeV/nucleon ($52\,$MeV/nucleon) are plotted as black (red) symbols. Only statistical errors are shown.}
   \label{fig:imbratio_QPb_bred}
  \end{figure}
  By exploiting the information provided by AMD+\textsc{Gemini++} on the correlation between $p_\text{red}$ and $b_\text{red}$ we can rescale on average the $x$-axis of Fig.~\ref{fig:imbratio_QPb_pred} into a $b_\text{red}$ axis. For each $R(\langle N/Z\rangle)$ branch, the $p_\text{red}$ vs $b_\text{red}$ correlation specifically obtained for the QP breakup channel in the corresponding asymmetric system at each beam energy has been used for the purpose; however, we point out that the behavior of such correlations is very similar to that found for the evaporative channel \cite{Ciampi2022}. The resulting $R(\langle N/Z\rangle)$ plots as a function of $b_\text{red}$ are shown in Fig.~\ref{fig:imbratio_QPb_bred} (using the same symbol and color code as before). At the cost of becoming model dependent, the plots in Fig.~\ref{fig:imbratio_QPb_bred} allow for a clearer comparison between the results at the two energies. As already found for the main binary exit channel \cite{Ciampi2022}, also in the case of QP breakup we find that a higher degree of equilibration is achieved at $32\,$MeV/nucleon than at $52\,$MeV/nucleon, a result in line with the literature \cite{Johnston1996} that we previously ascribed to the shorter interaction timescales in the latter case. The variation with the beam energy is more evident for the lower branch ($^{58}$Ni+$^{64}$Ni): the origin of such asymmetric behavior with respect to $R(\langle N/Z\rangle)=0$, possibly related to some statistical deexcitation effects not completely bypassed by the isospin transport ratio, is being investigated. In further studies, we will also consider the possibility that this effect could be partly caused by the fact that the total mass of the four systems is not equal.
  
 \subsection{Comparison between QP evaporation and breakup channels}
 \label{ssec:QPr-QPb_comparison}
 
 \begin{figure*}
  \centering
  \includegraphics[width=0.75\columnwidth]{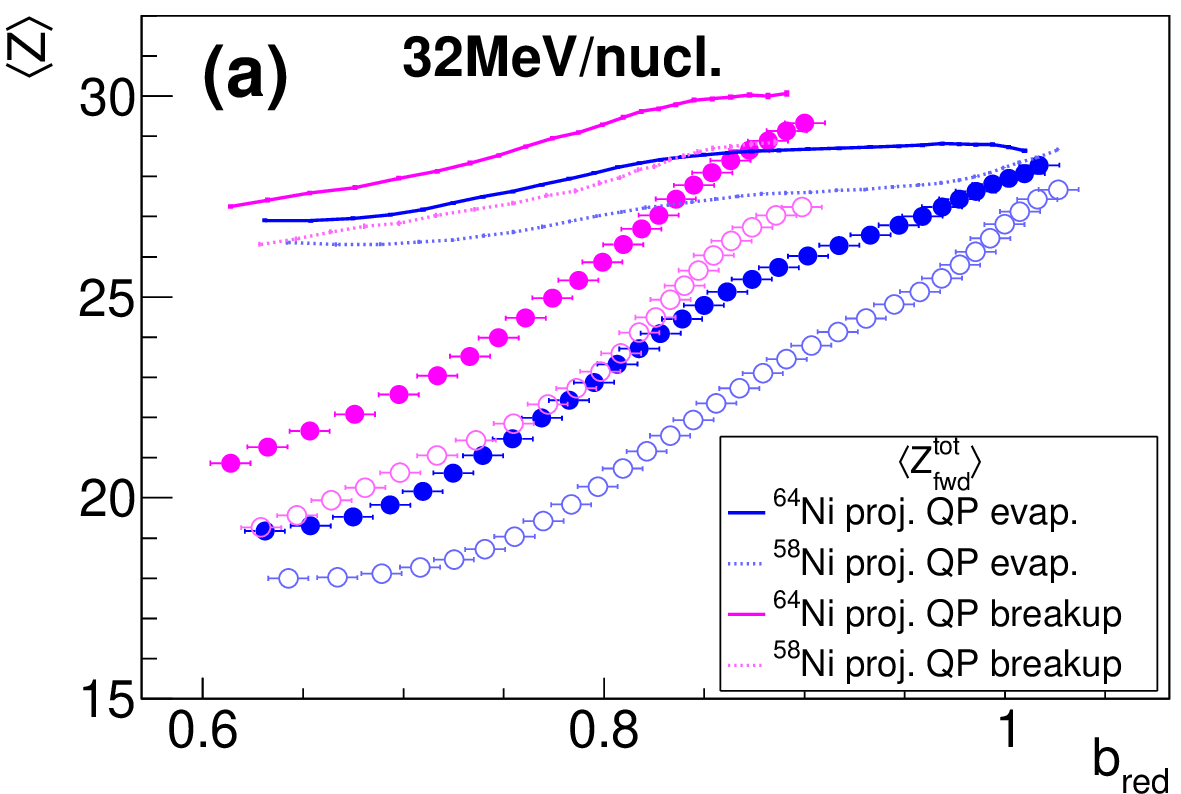}\qquad\qquad
  \includegraphics[width=0.75\columnwidth]{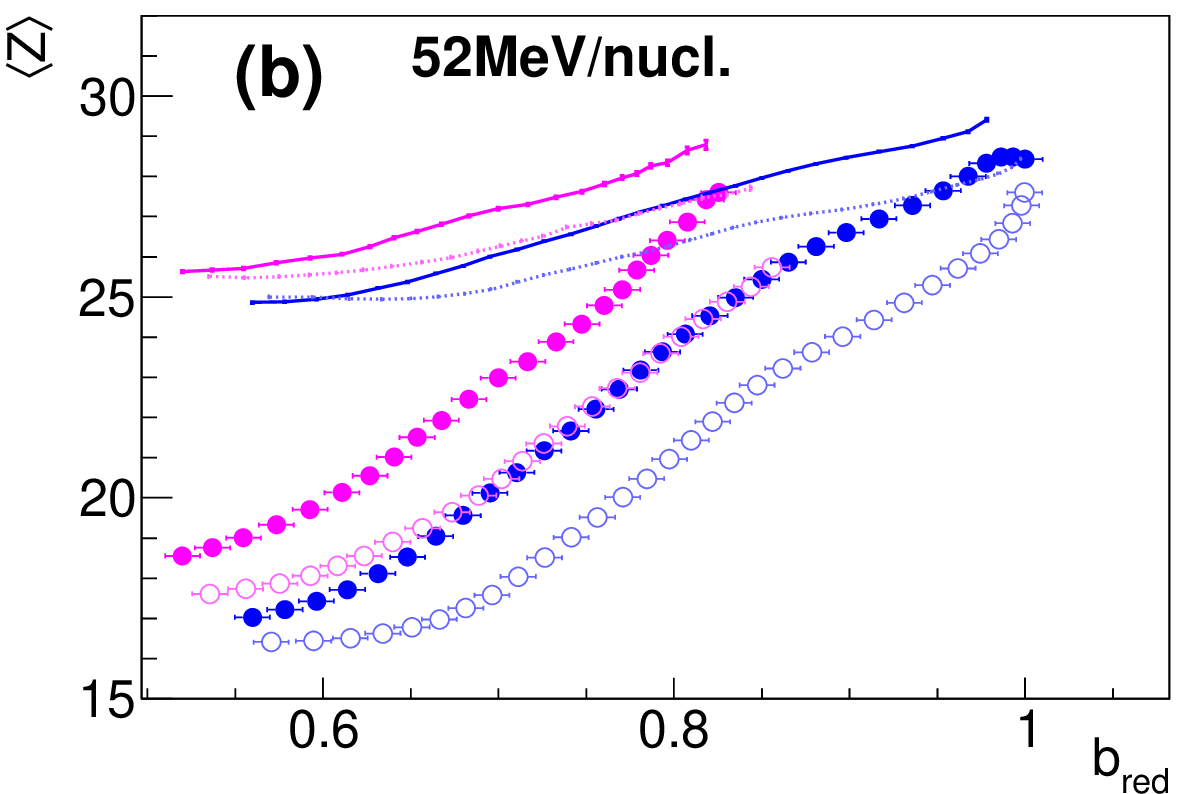}\\~\\
  \includegraphics[width=0.75\columnwidth]{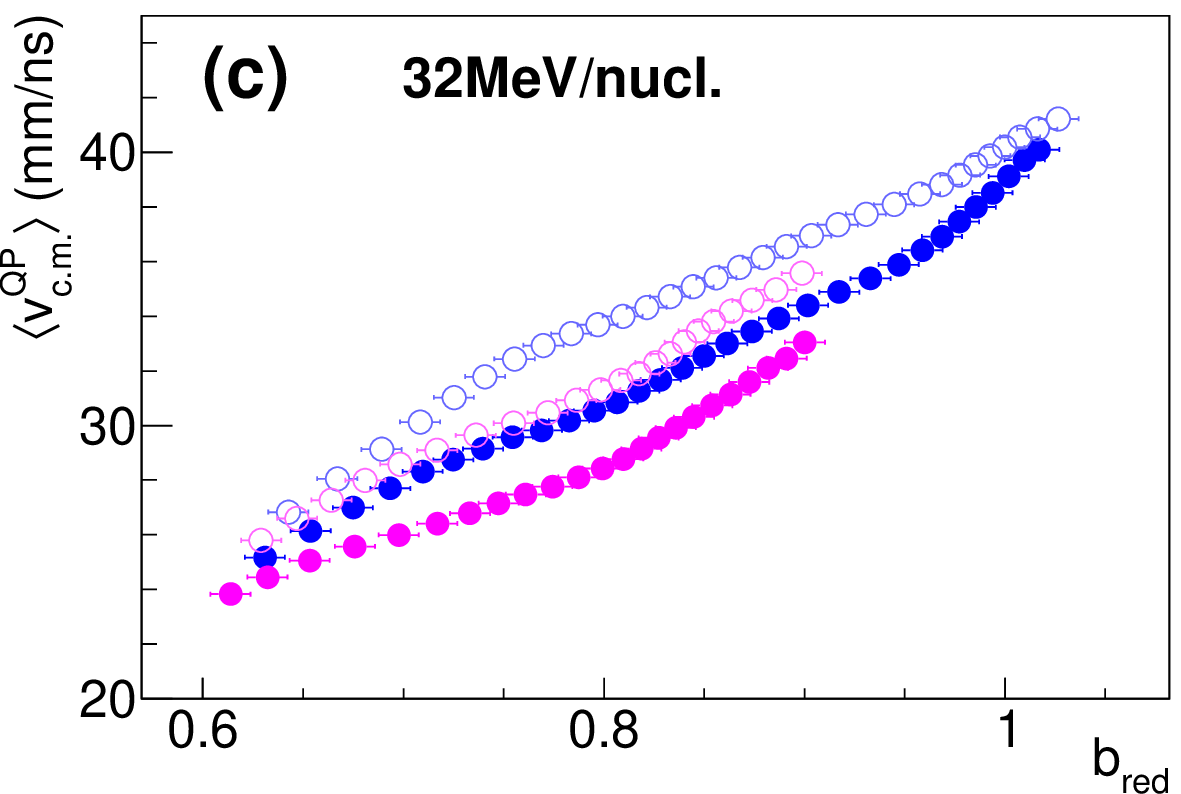}\qquad\qquad
  \includegraphics[width=0.75\columnwidth]{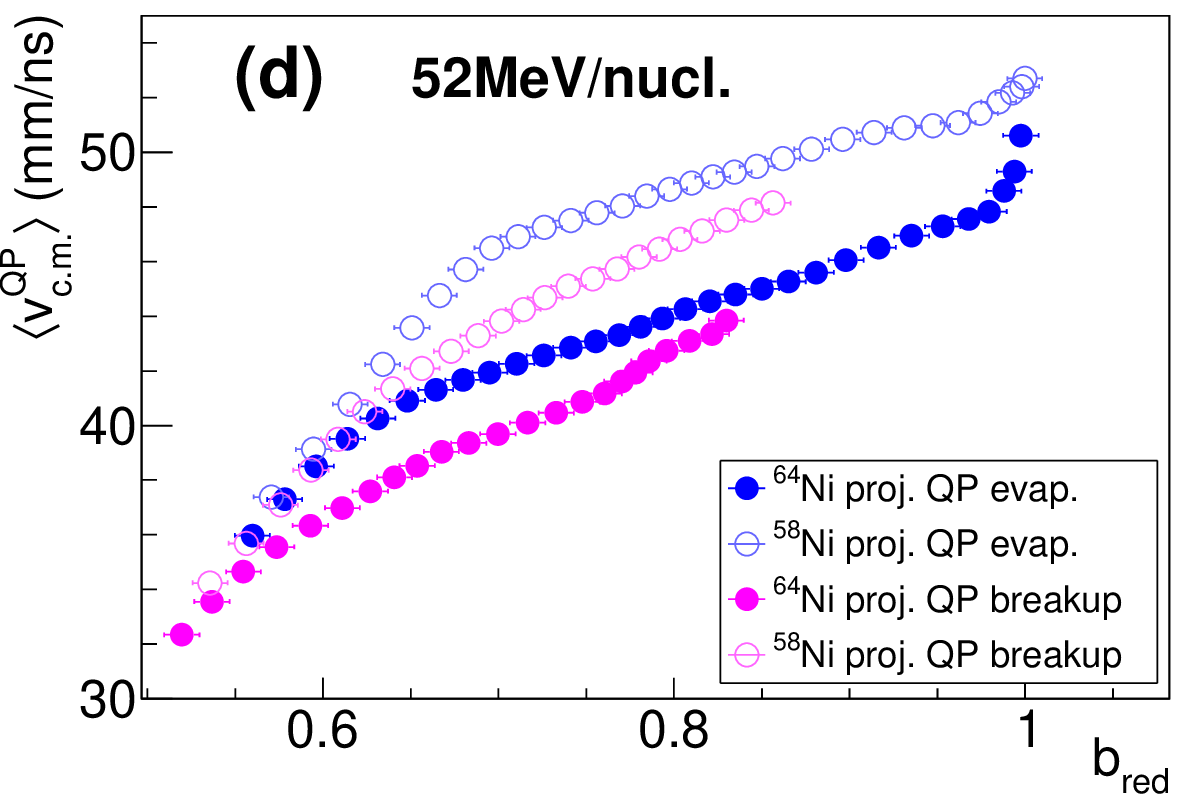}\\
  \caption{Comparison between the average charge [(a),~(b)] and velocity [(c),~(d)] of the QP remnant in the QP evaporation channel (in blue) and those of the reconstructed QP in the QP breakup channel (in magenta), for the reactions at $32\,$MeV/nucleon [(a),~(c)] and at $52\,$MeV/nucleon [(b),~(d)]. The plots are shown as a function of $b_\text{red}$, obtained by rescaling the $p_\text{red}$ order variable. At this level of investigation, the results obtained for the reactions induced by the same projectile are comparable and plotted only once, with solid (open) circles for the $^{64}$Ni ($^{58}$Ni) projectile; the $\langle Z^\text{tot}_\text{fwd}\rangle$ (see text) is plotted with continuous (dashed) lines.}
  \label{fig:Zmed_vcmmed_QPrQPb}
 \end{figure*}
 Further information on the reaction dynamics in semiperipheral collisions could be inferred by comparing the main properties and observations found for the two most populated reaction outputs, namely the QP evaporation channel \cite{Ciampi2022} and the QP breakup channel. We recall that the former exit channel has been selected by requiring the presence of only one forward emitted heavy fragment with $Z\geq15$, only accompanied by LCPs and, at most, IMFs. Some basic differences are already evident in the measured general properties of the QP residue and of the reconstructed QP, that we can compare still aware of the different reaction paths leading to those two channels. As an example, in Fig.~\ref{fig:Zmed_vcmmed_QPrQPb} we show the average charge $\langle Z\rangle$ and c.m.~velocity $\langle v \rangle$ of the QP remnant (blue) or reconstructed (magenta) as a function of $b_\text{red}$, for the reactions at both bombarding energies. Note that for the plots of Fig.~\ref{fig:Zmed_vcmmed_QPrQPb}, for a clearer comparison, we removed the mass identification requirement of the QP fragments, since it would affect differently the two reaction channels, especially for more peripheral collisions (the heaviest QP remnants would be indeed cut out due to the mass identification thresholds); such condition is however always imposed for the isospin analysis. We observe that, on average, the fissioning QP tends to be in all cases slightly heavier (with a difference of about 2-3 charge units) and slower than the non-fissioning one.
 It must be noted that the primary fragments produced in the two channels may evolve differently in the statistical deexcitation phase. As an example, we can assume that in the breakup channel, part of the original excitation energy is dissipated via the fission process itself, resulting in a shorter deexcitation path and thus in relatively heavier residues of the QP breakup fragments. We find confirmation of this consideration in the lower average multiplicity of light products in coincidence with the breakup fragments with respect to the other channel (not shown). In order to take into account the effect, in Figs.~\ref{fig:Zmed_vcmmed_QPrQPb}(a),(b) we also plot as blue and magenta lines the average total charge detected in the forward hemisphere in the c.m.~reference frame $\langle Z^\text{tot}_\text{fwd}\rangle$. A larger total charge is found in the breakup channel, even exceeding the original projectile charge for less central collisions at the lower beam energy. The difference between the two channels, albeit reduced down to only about 1 charge unit, is still present, possibly indicating that the two outputs correspond to different dynamical evolutions of the interacting system already prior to the deexcitation phase.
 
 \begin{figure*}
  \centering
  \includegraphics[width=0.9\columnwidth]{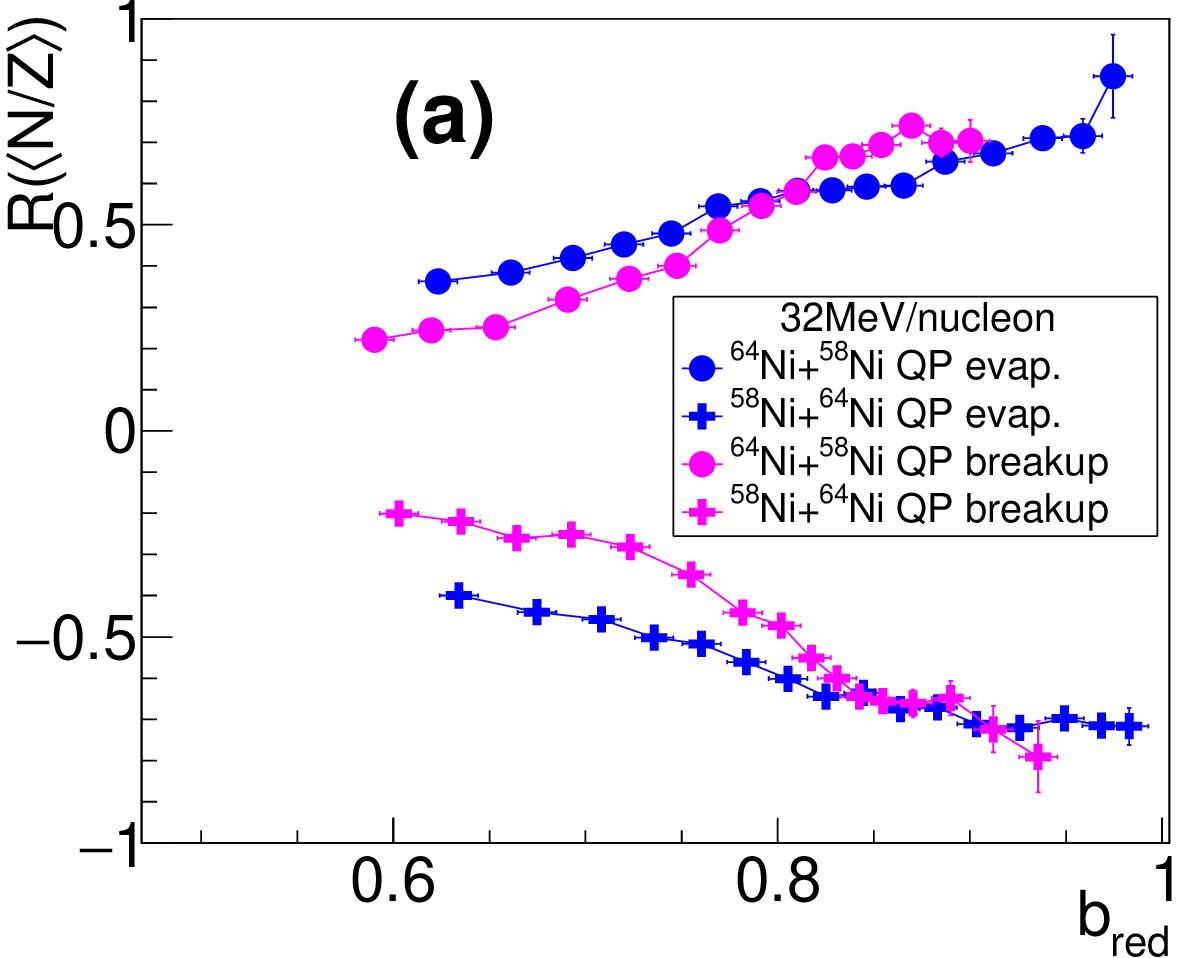}$\qquad\qquad$
  \includegraphics[width=0.9\columnwidth]{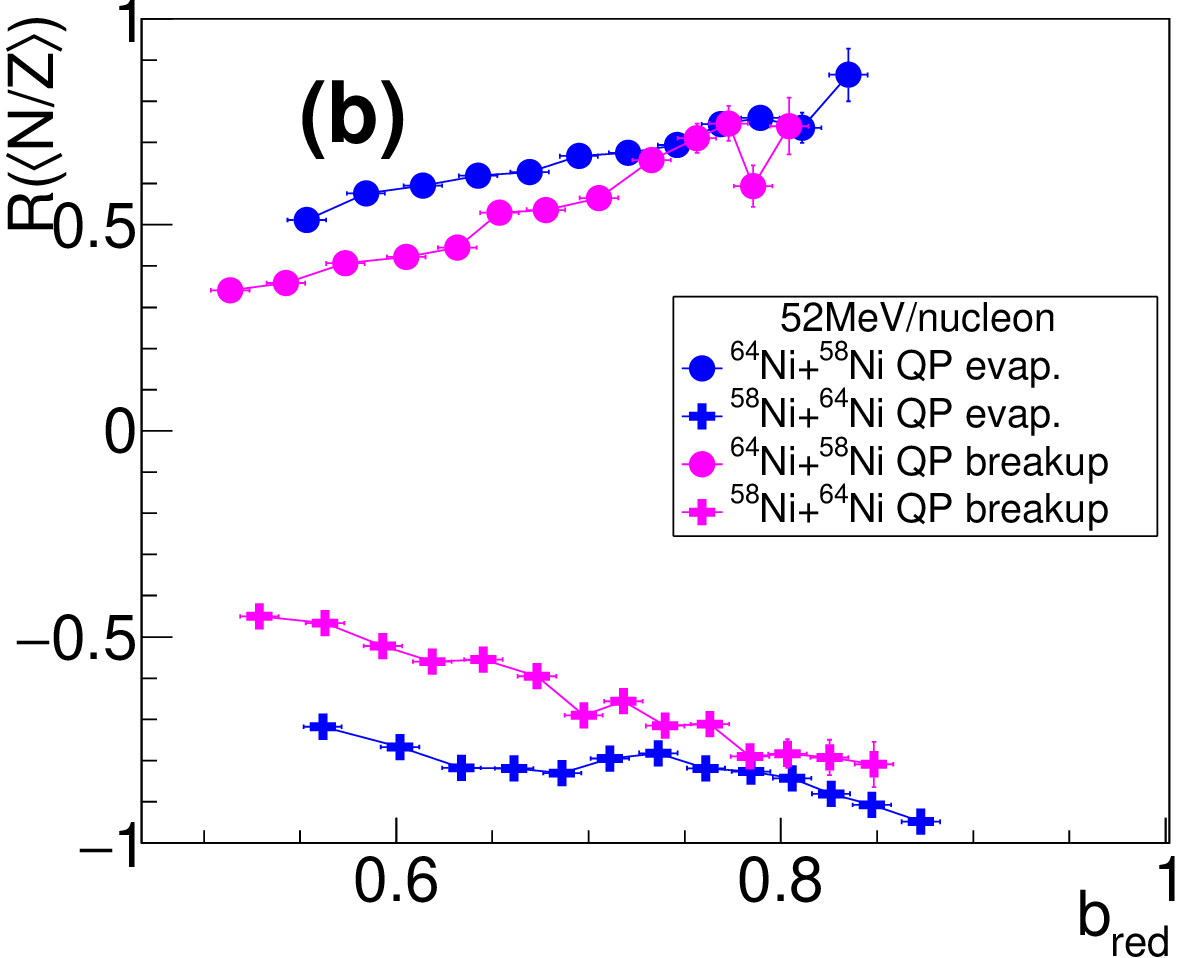}  
  \caption{Comparison between the experimental results in the QP evaporation and QP breakup channel selections of the isospin transport ratio calculated with the $\langle N/Z\rangle$ of the QP (residue or reconstructed, respectively), reported as a function of $b_\text{red}$, for the reactions at (a) $32\,$MeV/nucleon and (b) at $52\,$MeV/nucleon. Only statistical errors are shown.}
  \label{fig:imbratio_QPrQPb_bred}
 \end{figure*}
 We further explore this point by exploiting the information provided by the purely dynamical effect of isospin diffusion. In Fig.~\ref{fig:imbratio_QPrQPb_bred} we present the comparison between the isospin transport ratios obtained for the two selected exit channels (using the same color code as before), reported as a function of $b_\text{red}$,
 by rescaling the $p_\text{red}$ variable according to the $b_\text{red}$ vs $p_\text{red}$ correlation for the respective channel. 
 In both plots, a quite general observation arises: the breakup channel generally shows a stronger trend towards isospin equilibration with respect to the evaporation channel for the same asymmetric reaction. 
 At both energies, the difference between the two channels is larger between the lower branches of $R(\langle N/Z\rangle)$, i.e., for the reaction $^{58}$Ni+$^{64}$Ni (solid crosses), but it is still evident also for the other mixed system, $^{64}$Ni+$^{58}$Ni (solid circles), even though at $32\,$MeV/nucleon it is present only for less peripheral collisions. 
  
 It is worth mentioning that a comparison between the isospin transport ratio in the breakup and in the evaporation channel has been done by the FAZIA collaboration also in a previous experiment \cite{Camaiani2021} carried out in 2015, right after the FAZIA R\&D phase, with a 4 blocks configuration covering the forward angles up to $\theta_\text{lab}\sim8$\textdegree. In that work, the asymmetric reaction $^{48}$Ca+$^{40}$Ca at 35~MeV/nucleon has been compared to the two corresponding symmetric reactions, thus allowing to build only the upper branch of $R(\langle N/Z\rangle)$: the comparison between the two exit channels could not reveal any significant difference between them. However, we verified that the AMD+\textsc{Gemini++} simulations of Ca+Ca systems predict the aforementioned effect, with a quite limited difference for the upper branch and a more evident gap for the lower one, similarly to the behavior experimentally observed also for the Ni+Ni systems. In the previous FAZIA experiment, both the availability of only the $^{48}$Ca+$^{40}$Ca mixed projectile-target combination and the lower angular coverage may have somewhat blurred the observation of the different isospin equilibration in the two selected output channels: this evidences how having access to the full set of reactions and a large detection solid angle are needed in order to provide the most complete information on the isospin equilibration process when the isospin transport ratio technique is used.
 
 The new experimental evidence of Fig.~\ref{fig:imbratio_QPrQPb_bred} deserves deeper investigation: the QP breakup channel seems to indirectly select, at the end of the secondary phase, a set of events where a more prominent role has been played by the isospin diffusion between projectile and target, resulting in a more isospin equilibrated QP. This may be due to the indirect selection of events in the tail of the distribution of some parameters more related to the reaction dynamics. 
 In fact, due to the intrinsic fluctuations associated with the involved dissipative processes, even for the same entrance channel conditions (e.g., same system, beam energy, impact parameter), the key quantities describing the properties of the interacting system can assume a whole range of values. Such dynamical parameters cannot be directly accessed from an experimental point of view. However, in this framework, we attempted to give an interpretation of the different trends towards isospin equilibration observed in the two selected exit channels, basing on mostly semiclassical arguments. As a first approximation, a different degree of relaxation of the initial isospin imbalance can be related to different projectile-target contact times during the interaction phase. A longer contact time during the reseparation of QP and QT could in turn lead to a more pronounced deformation of the interacting system, with a progressively more elongated neck structure connecting QP and QT. At the time of their reseparation, the primary QP and/or the primary QT (or both of them) could hence inherit a stronger deformation, possibly due to a partial reabsorption of the neck structure, and such deformation together with the angular momentum of the product could in the end result in a QP/QT breakup event.
 We point out that the proposed picture is quite a naive interpretation, which does not take into account other crucial quantities, such as the nuclear density range explored in the contact area, that surely play a key role in the process. Nevertheless, it could explain, at least partially,  why a stronger isospin equilibration is found for the class of events we identified as QP breakup. In order to verify this hypothesis we rely on the AMD model, as discussed in the following section.
 
 \section{Discussion and interpretation}
 \label{ssec:contact_time}
  We can exploit the information provided by the AMD simulations to investigate possible differences between the dynamical scenarios in the primary events associated with the QP evaporation and QP breakup outputs. In doing this, for clarity's sake, we have to report that the experimental neutron content of the secondary fragments is not well reproduced by the filtered simulations, as already pointed out in the literature for the same combination of codes \cite{Piantelli2020}, with a very strong role of the employed afterburner. However, the statistical code contribution is largely reduced by exploiting the isospin transport ratio, and the property we are interested to track is the contact time.
  
  In order to exclude any possible role of the apparatus response, in this phase we analyzed the unfiltered simulated datasets, which also allow to access the QT phase space. For this purpose, the two channel selection criteria (still applied at the end of the afterburner stage) have been modified and adapted to the $4\pi$ solid angle geometry.
  For the QP evaporation channel we require the presence of two heavy fragments with $Z\geq15$, one in the forward hemisphere in the c.m.~reference frame, the other in the backward one, with only LCPs and IMFs in coincidence. The main difference between this channel selection in $4\pi$ and the corresponding one in the experimental data (or in the filtered simulations) lies in the complete removal of the QT breakup events with both fission fragments below the detection threshold.
  Concerning the QP breakup selection, we proceed slightly differently than in the experimental case, by selecting the events with only two forward emitted $Z\geq5$ fragments in the c.m.~reference frame, whose charge number sum is at least 15; however, we have checked that this condition selects a distribution of $\theta_\text{rel}$ between the two heavy fragments which largely corresponds to the one evidenced with the red cut in Fig.~\ref{fig:thetarel_vrel}.
  We do not put conditions on the heavy fragment multiplicity in the backward hemisphere, so that the selected QP breakup class also includes double breakup events, which however constitute only a minor contribution, mostly at lower $b_\text{red}$ values.
  Moreover, we also label the ``complementary" subclass (i.e., with two backward emitted heavy fragments with $Z_\text{rec}\geq15$ and only one forward) as QT-only breakup events.
  Adopting these selection criteria, also in the unfiltered AMD+\textsc{Gemini++} model predictions the different trend towards isospin equilibration of the two exit channels (evaporation and breakup) is observed.

  In order to extract the contact time information we exploit a method based on a tool already presented in Ref.~\cite{Piantelli2020}: for each primary event, the AMD fragment reconstruction algorithm is applied every 20~(10)~fm/$c$ for the reactions at 32~(52)~MeV/nucleon, so that a picture of the event at each timestep from 0 to 500~fm/$c$ is stored. Next, each event is read from the start, looking for the first timestep at which only one heavy fragment is found, labelled as $t_\text{stick}$, and then for the first timestep after $t_\text{stick}$ at which at least two heavy fragments emerge again, labelled as $t_\text{DIC}$ as in Ref.~\cite{Piantelli2020}. The contact time between projectile and target is therefore calculated as the difference $t_\text{cont}=t_\text{DIC}-t_\text{stick}$.
  
\begin{figure}
   \centering
   \includegraphics[width=0.78\columnwidth]{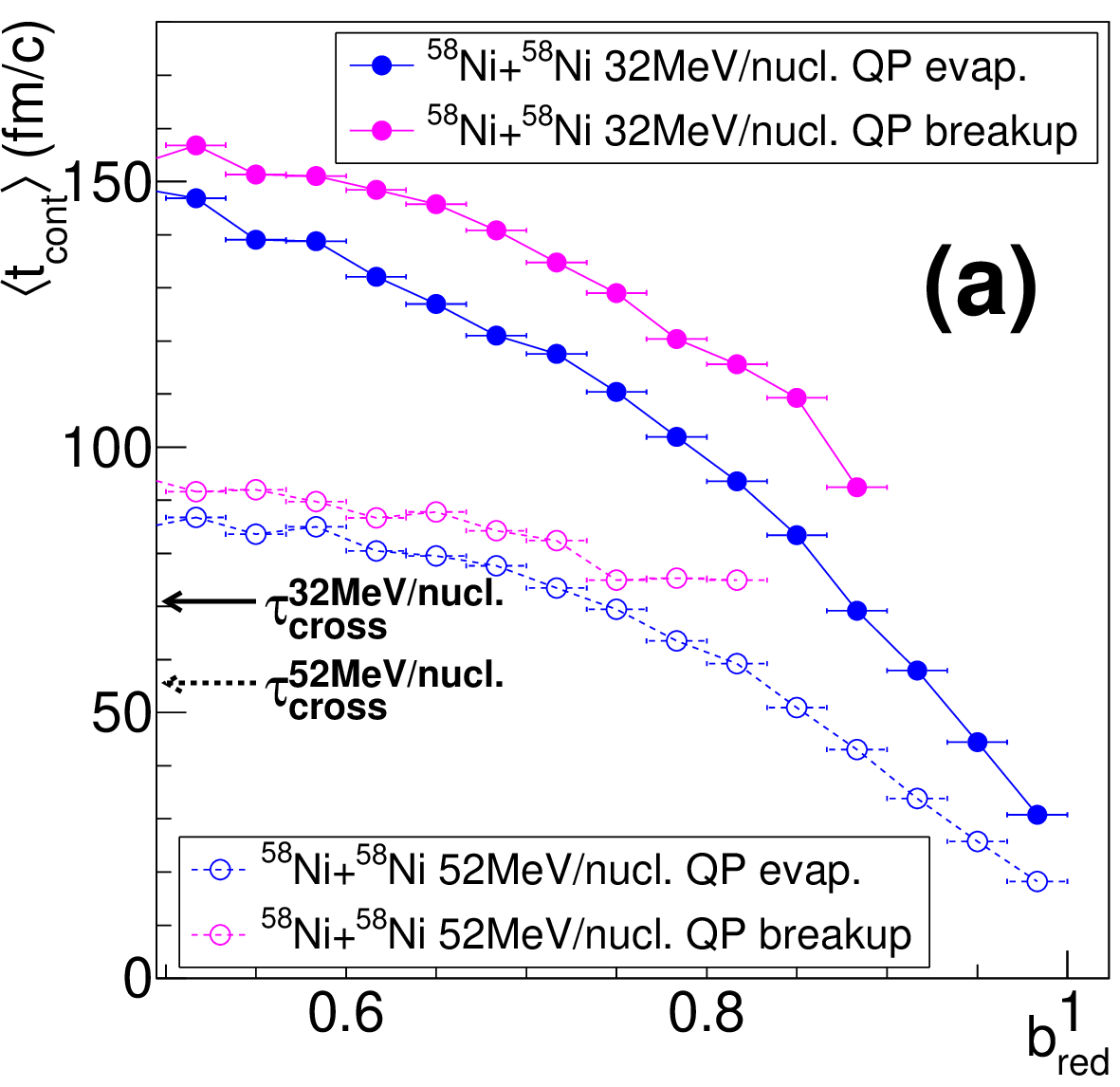}\\
   \includegraphics[width=0.78\columnwidth]{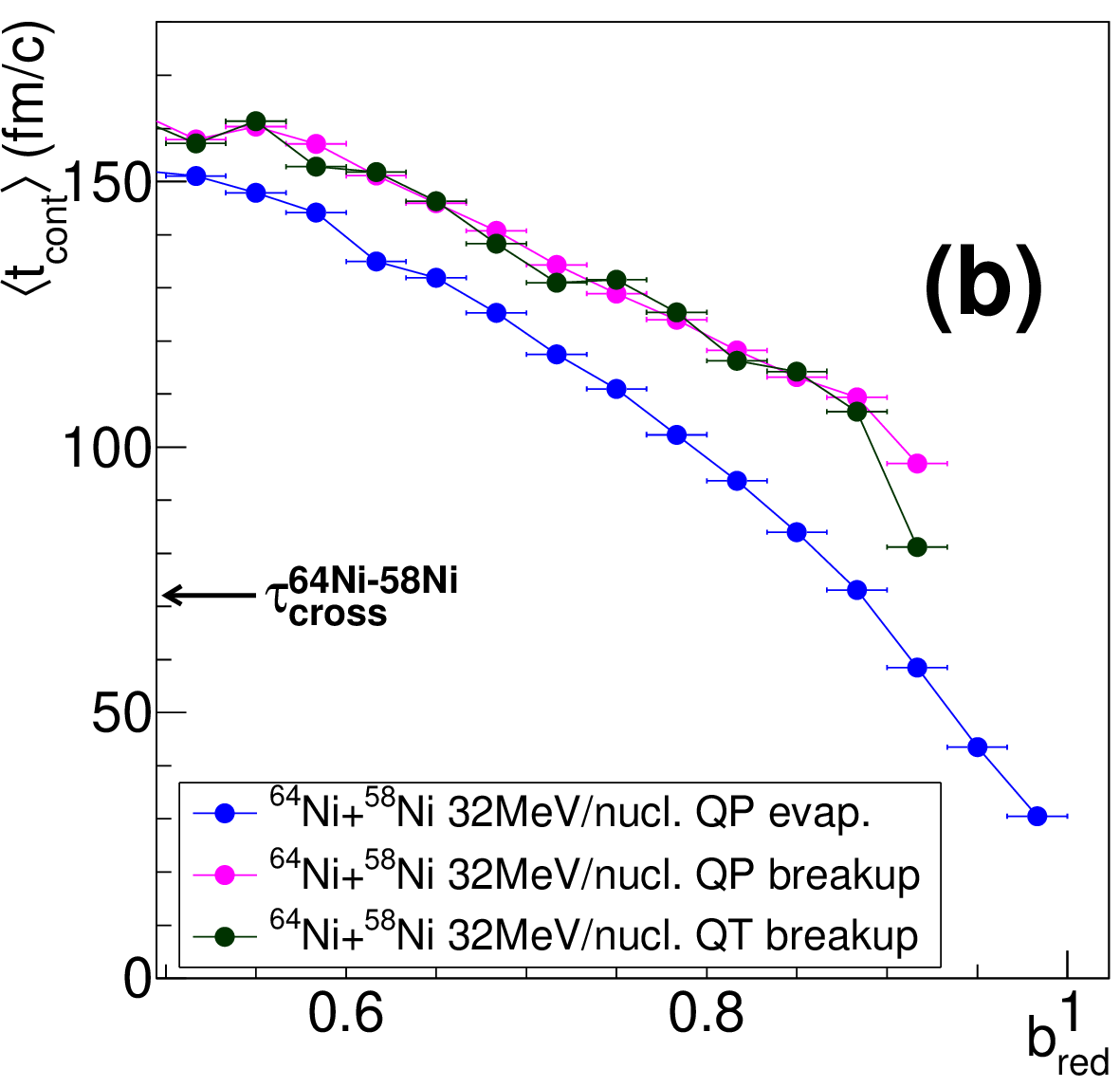}\\
   \includegraphics[width=0.78\columnwidth]{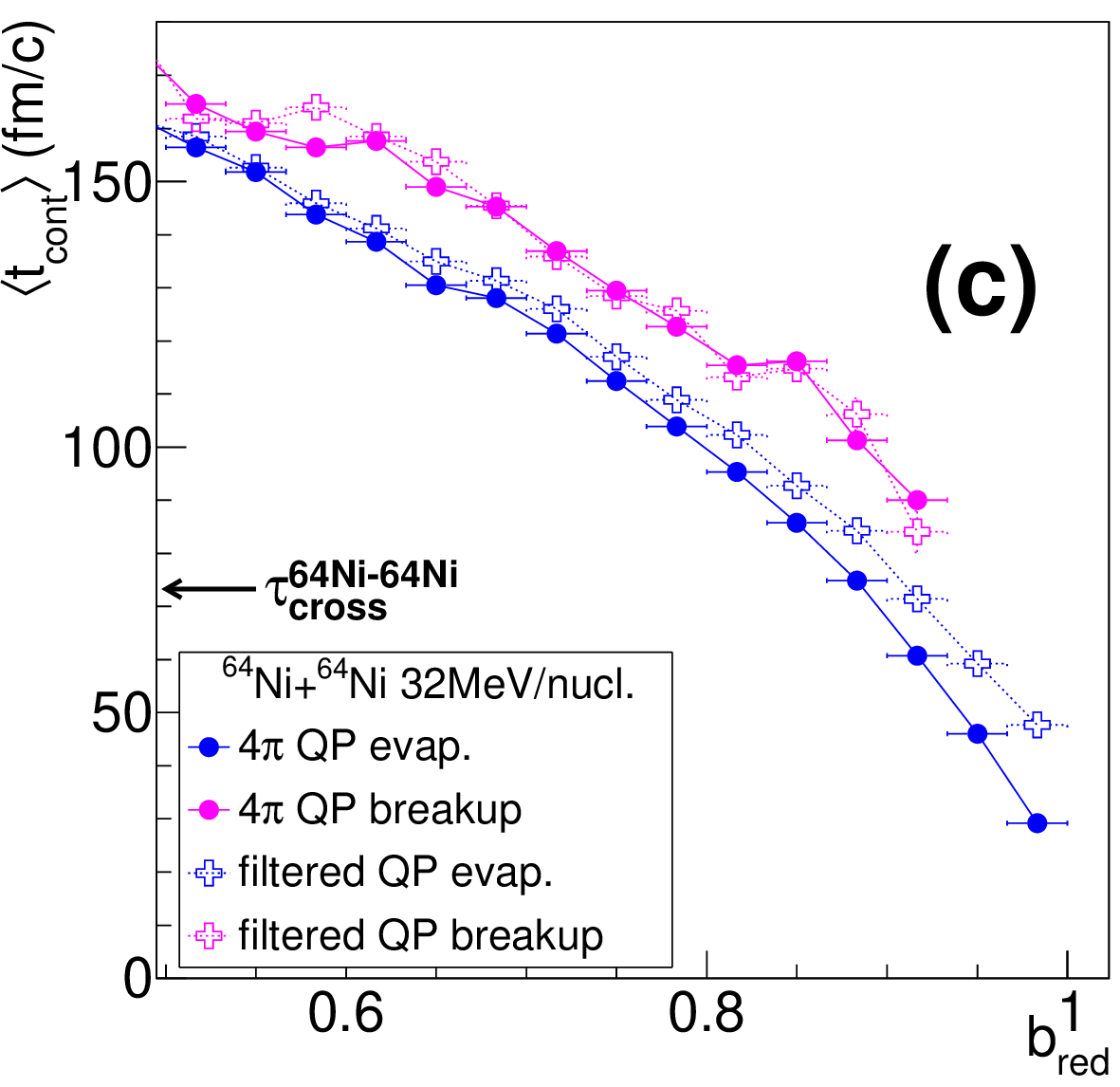}
   \caption{Model results: average contact times $\langle t_\text{cont}\rangle$ as a function of $b_\text{red}$ extracted from the $4\pi$ AMD+\textsc{Gemini++} simulations for the different selected output channels for (a) the reaction $^{58}$Ni+$^{58}$Ni at 32 and 52~MeV/nucleon, (b) the reaction $^{64}$Ni+$^{58}$Ni at 32~MeV/nucleon and (c) the reaction $^{64}$Ni+$^{64}$Ni at 32~MeV/nucleon, the latter both before ($4\pi$) and after the filtering of the simulated dataset. The calculated crossing times $\tau_\text{cross}$ for the respective systems are indicated with the black arrows on the left side of each plot.}
   \label{fig:tempi_contatto}
  \end{figure}

  The average contact times $\langle t_\text{cont}\rangle$ thus obtained for the different systems are shown in Fig.~\ref{fig:tempi_contatto} as a function of the reaction centrality for the exit channels selected at the end of the secondary phase in the $4\pi$ analysis as described above, using the same color code of the previous plots for QP evaporation (blue) and QP breakup (magenta). Each panel refers to a different system. As a reference, on each plot we also indicate with a black arrow the nuclear crossing time for the different reactions, calculated assuming a trivial central collision geometry as $\tau_\text{cross}= 2R_0(A_p^{1/3}+A_t^{1/3})/ v_\text{beam}^\text{lab}$, where $R_0=1.2\,$fm and $A_p$ and $A_t$ are the mass numbers of projectile and target. 
  The latter evaluation has been included in order to show that the $t_\text{cont}$ values extracted from the AMD calculation behave quite reasonably, at least for the $b_\text{red}$ interval under investigation, in which the end of the contact phase $t_\text{DIC}$ can be defined sufficiently clearly. We observe that for the binary output (in blue) of a given reaction, $t_\text{cont}\approx \tau_\text{cross}$ for the most peripheral collisions (i.e., with $b_\text{red}\sim0.9$). With increasing centrality, they also increase monotonously to values larger than $\tau_\text{cross}$, as we can expect for a process which is slowed down by the nuclear interaction itself. Moreover, we point out that the $\langle t_\text{cont}\rangle$ scales correctly with the beam energy (see Fig.~\ref{fig:tempi_contatto}(a)) and with the size of the system (compare the three panels of Fig.~\ref{fig:tempi_contatto}): interestingly, despite the very limited interaction timescale variation that we could expect due to the different system size (refer, e.g., to the $\tau_\text{cross}$ indications), the corresponding small $\langle t_\text{cont}\rangle$ variation can be evidenced.
  These observations support the reliability and sensitivity of the average contact time information, within the framework of a model calculation.
  Moving to the comparison between the two investigated output channels, a quite general observation arises from all the inspected systems: according to AMD+\textsc{Gemini++}, slightly longer projectile-target contact times seem to be on average associated with the breakup channel with respect to the evaporative one, consistently with the simplified hypothesis that we have hereby proposed. 
  
  Going into more detail, in Fig.~\ref{fig:tempi_contatto}(a) we show the results for the same system $^{58}$Ni+$^{58}$Ni at the two beam energies 32~MeV/nucleon (solid markers, solid line) and 52~MeV/nucleon (open markers, dashed line). The systematic difference between the two channels is rather evident at the lower beam energy, with $\Delta \langle t_\text{cont}\rangle$ ranging approximately between 15 and 20~fm/$c$. On the other hand, at 52~MeV/nucleon, such difference is considerably smaller, though still noticeable ($\sim5-10$~fm/$c$) within the overall shorter timescales involved.
  
  In Fig.~\ref{fig:tempi_contatto}(b) we present the average contact times for the asymmetric system $^{64}$Ni+$^{58}$Ni at 32~MeV/nucleon. Together with the results obtained in the QP evaporation and QP breakup channels, by exploiting the $4\pi$ analysis, in this plot we also show the QT breakup channel, in black: in fact, following the proposed interpretation, a longer contact time could equivalently result in a more pronounced deformation, and hence a breakup, of the QP or the QT, with no preferential outcome. Consistently, we obtain a comparable $\langle t_\text{cont}\rangle$ for the two breakup channels, again longer than that characteristic of the evaporative output. In this $4\pi$ analysis, carried out with symmetric channel selection criteria (with the only exception of double breakup events), the same evaluations for the other asymmetric system $^{58}$Ni+$^{64}$Ni can be inferred from the plot in Fig.~\ref{fig:tempi_contatto}(b).
  
  The result for the heaviest system of the dataset $^{64}$Ni+$^{64}$Ni at the lower beam energy is shown in Fig.~\ref{fig:tempi_contatto}(c): here, we also compare the $\langle t_\text{cont}\rangle$ vs $b_\text{red}$ extracted according to the event selection performed on the unfiltered AMD+\textsc{Gemini++} simulations (solid circles, solid line) with what is obtained after the filtering procedure (open crosses, dashed line). Besides the considerations set out above on the $\Delta \langle t_\text{cont}\rangle$ between the two channels, which are confirmed both before and after the filtering, we notice that the experimental acceptance does not play a sizeable role on the average contact time in the QP breakup selection, while it seems to select slightly longer interactions in the QP evaporation class, especially at larger $b_\text{red}$. Two contributions may be responsible of such variation found for the filtered dataset. The first one, with a minor role and mostly limited to less peripheral events, is due to the possible contamination of the QP evaporation class with QT breakup events where both QT fission fragments are undetected: as seen from Fig.~\ref{fig:tempi_contatto}(b), such events would contribute with the longer $t_\text{cont}$ of a breakup event. The second contribution, which is more important and characteristic of more peripheral events, is due to the mass identification limit for heavy products: heavier QP remnants, namely those mostly affected by such experimental cut, are in fact produced in less dissipative collisions, likely related to shorter interaction times, that will be hence selectively removed from the dataset. As a consequence of such experimental bias, we can also expect an alteration of the isospin transport ratio for $b_\text{red}\sim1$, since the removed events are also plausibly related to a weaker isospin equilibration.

\section{Summary and conclusions}
 In this work we presented an analysis of experimental data for the four systems $^{58,64}$Ni+$^{58,64}$Ni at the two beam energies 32 and 52~MeV/nucleon, acquired in the first campaign of the coupled INDRA-FAZIA apparatus \cite{Ciampi2022}, focusing on the ternary exit channel of semiperipheral and peripheral collisions, identified as the one related to QP fission events. 
 The behavior of the characteristic features of the fission process, namely the mass asymmetry and the orientation of the QP split, is consistent with a fast dynamical breakup \cite{Casini1993,Stefanini1995,Piantelli2020,Bocage2000,DeFilippo2005}: indeed, we observed a clear correlation between larger mass asymmetries and spatial configurations more aligned to the QP-QT axis, with the light fission fragment emitted towards the c.m.~of the reaction.
 The investigated reactions have also been simulated by means of the AMD dynamical model, coupled to \textsc{Gemini++} as afterburner and filtered according to the experimental acceptance: the main features of the simulated reaction products show a quite good agreement with the experimental data. In the framework of the model predictions, a large majority of the fission events are present already at the end of the dynamical code calculation (i.e., they are produced before 500~fm/$c$): we can therefore quite safely state that the ternary exit channel that we select is mostly populated by QP dynamical fission, or QP breakup, events. 

 Moving to the analysis of the isospin content of the QP reconstructed from the two fission fragments, thanks to the availability of data for both the asymmetric and symmetric combinations of $^{58}$Ni and $^{64}$Ni, we have been able to fully exploit the technique of the isospin transport ratio \cite{Rami2000} (calculated with the average neutron-to-proton ratio of the reconstructed QP) to highlight the trend towards the relaxation of the isospin imbalance in the two asymmetric systems. A stronger isospin equilibration is achieved at the lower beam energy, as already observed for the same dataset in Ref.~\cite{Ciampi2022} for the evaporative channel: such finding can be qualitatively explained by the longer interaction timescales at lower energy.
 Quite interestingly, however, by comparing the isospin transport ratios built in the two selected channels, we also observed a more pronounced trend towards isospin equilibration for the reconstructed QP in the breakup channel than for the QP remnant in the binary output, for both beam energies. 

 According to the experimental observations hereby shown, the QP breakup channel seems to select, for a given centrality bin, a subset of events where a more effective isospin diffusion process took place, possibly due to the indirect selection of reactions characterized by a set of dynamical parameters closer to the edges of their typical distributions. In an attempt of explaining this experimental result, we proposed a simplified interpretation, that relates QP/QT breakups to events in which a prolonged contact phase between projectile and target induces a more pronounced deformation of the primary QP/QT, which in turn facilitates their fission.
 Since the same effect of stronger isospin equilibration is found also in the QP breakup selection in the simulated dataset, we exploited the AMD calculations to verify whether our hypothesis could be supported by the model predictions. By running the AMD fragment reconstruction algorithm every 20~fm/$c$ timestep (10~fm/$c$ for the reactions at the higher beam energy) \cite{Piantelli2020} we were able to extract the contact time between projectile and target for each event. 
 The comparison of the average contact time as a function of the reaction centrality obtained for the two reaction channels showed that the breakup channel indeed follows averagely longer projectile-target interactions.
 The contact time difference between the two exit channels is rather small (15-20~fm/$c$ at 32~MeV/nucleon, 5-10~fm/$c$ at 52~MeV/nucleon), but it is systematically present within the reaction centrality interval under investigation. We point out that such gap may not be sufficient to fully explain the difference in the two trends towards isospin equilibration which has been observed experimentally: other key quantities may play an equally important role.

 In conclusion, in this work, the optimum performance of the INDRA-FAZIA apparatus has allowed to experimentally observe various features related to the QP dynamical breakup that we believe add valuable information for a comprehensive view of the process. Further experimental investigations, also taking into account different reaction systems, in parallel with an in-depth comparison with model predictions, are needed for the most complete understanding of the phenomenon, in view of exploiting it to the fullest to help constrain fundamental ingredients contained in reaction models.

 \begin{acknowledgments}
 This work was in part supported by the National Research Foundation of Korea (NRF; Grant No.~2018R1A5A1025563), by the Institute of Basic Science, Republic of Korea (IBS; Grant No.~IBS-R031-D1) and by the Spanish Ministerio de Economía y Empresa (PGC2018-096994-B-C22).
 We acknowledge support from Région Normandie under Réseau d’Intérêt Normand FIDNEOS (RIN/FIDNEOS).
 Many thanks are due to the accelerator staff of GANIL for delivering a very good quality beam and to the technical staff for the continuous support.
\end{acknowledgments}

 \bibliographystyle{apsrev4-2.bst}
 
\providecommand{\noopsort}[1]{}

\end{document}